\documentclass[10pt,aps,prd,showpacs,nofootinbib,twocolumn,superscriptaddress]{revtex4-2}

\usepackage{amssymb,amsmath,amsfonts}
\usepackage[usenames]{xcolor}
\usepackage[pdftex]{graphicx}
\usepackage[english]{babel}
\usepackage{amsmath,amssymb}
\usepackage{multirow}
\usepackage{txfonts}
\usepackage[percent]{overpic}

\usepackage{lipsum}

\newcommand{\avgcs}[0]{\ensuremath{\langle c_s^2\rangle~}}

\begin{document}

\title{I -- Love -- \avgcs: Approximately universal relations for the average neutron star stiffness}

\author{Jayana A. Saes}
\email{jayanaa2@illinois.edu}
\affiliation{Illinois Center for Advanced Studies of the Universe, Department of Physics, University of Illinois at Urbana-Champaign, Urbana, IL 61801, USA}
\author{Raissa F.\ P.\ Mendes}
\email{rfpmendes@id.uff.br}
\affiliation{Instituto de F\'isica, Universidade Federal Fluminense, Niter\'oi, Rio de Janeiro, 24210-346, Brazil}
\affiliation{CBPF - Centro Brasileiro de Pesquisas F\'isicas, Rio de Janeiro, 22290-180, Brazil}
\author{Nicol\'as Yunes}
\email{nyunes@illinois.edu}
\affiliation{Illinois Center for Advanced Studies of the Universe, Department of Physics, University of Illinois at Urbana-Champaign, Urbana, IL 61801, USA}

\date{\today}

\begin{abstract}

The accurate observations of neutron stars have deepened our knowledge of both general relativity and the properties of nuclear physics at large densities. 
Relating observations to the microphysics that govern these stars can sometimes be aided by approximate universal relations. 
One such relation connects the ratio of the central pressure to the central energy density and the compactness of the star, and it has been found to be insensitive to realistic models for the equation of state to a $\sim 10\%$ level. 
In this paper, we clarify the meaning of the microscopic quantity appearing in this relation, which is reinterpreted as the average of the speed of sound squared in the interior of a star, \avgcs$\!$. 
The physical origin of the quasi-universality of the $\avgcs - C$ relation is then investigated. Using post-Minkowskian expansions, we find it to be linked to the Newtonian limit of the structure equations and the fact that the equations of state that describe NSs are relatively stiff. The same post-Minkowskian approach is also applied to the relations between \avgcs$\!$, the moment of inertia, and the tidal deformability of a neutron star, in which cases the same degree of universality is found across post-Minkowskian orders.
\end{abstract}


\maketitle

\allowdisplaybreaks[4] 

\section{Introduction}
\label{sec:intro}

Neutron stars (NSs) are highly compact astrophysical objects, in which matter exists at extremely dense levels, with their maximum density possibly reaching up to $\sim$10 times the saturation density $\rho_{sat} = 2.7\times 10^{14}$g/cm$^3$. This unique environment renders a NS an ideal laboratory for fundamental physics, encompassing tests of general relativity (GR) and investigations into the equation of state (EoS) of supra-nuclear matter.
To realize this potential, accurate measurements of NS observables are needed. 
Recent gravitational wave (GW) detections of binary NS mergers \cite{LIGOScientific:2017vwq,LIGOScientific:2018cki}, radio observations of pulsars \cite{Kramer:2006nb,Weisberg:2010zz, Fonseca:2014qla}, and pulse-profiling of X-ray data from the Neutron-Star-Interior Composition Explorer (NICER) \cite{Miller:2019cac, Riley:2019yda, Miller:2021qha, Riley:2021pdl}, are beginning to provide the first accurate measurements of NS radii, masses, and tidal deformabilities. 
Next-generation GW observatories such as the Einstein Telescope \cite{Maggiore:2019uih} and the Cosmic Explorer \cite{Evans:2021gyd}, as well as future X-ray missions, will allow an increase in abundance and precision of such measurements.

An important tool for the interpretation of NS observational data and its linkage to the microphysics governing the interior of these stars lies in approximate universal relations, i.e.,  EoS-insensitive relations between NS properties. One example is the I-Love-Q relations \cite{Yagi:2013awa}, which connect (a dimensionless version of) the moment of inertia, tidal deformability, and quadrupole moment to each other, and can be used to break degeneracies in GW parameter estimation. Another example is the binary Love relations \cite{Yagi:2015pkc, Yagi:2016qmr}, which connect (a dimensionless version of) the tidal deformabilities of two NSs in a binary to each other and can enable the measurement of individual tidal deformabilities from GW data. For a review of approximate universal relations, we refer the reader to Ref.~\cite{Yagi:2016bkt}.

\begin{figure*}
\includegraphics[width=\textwidth]{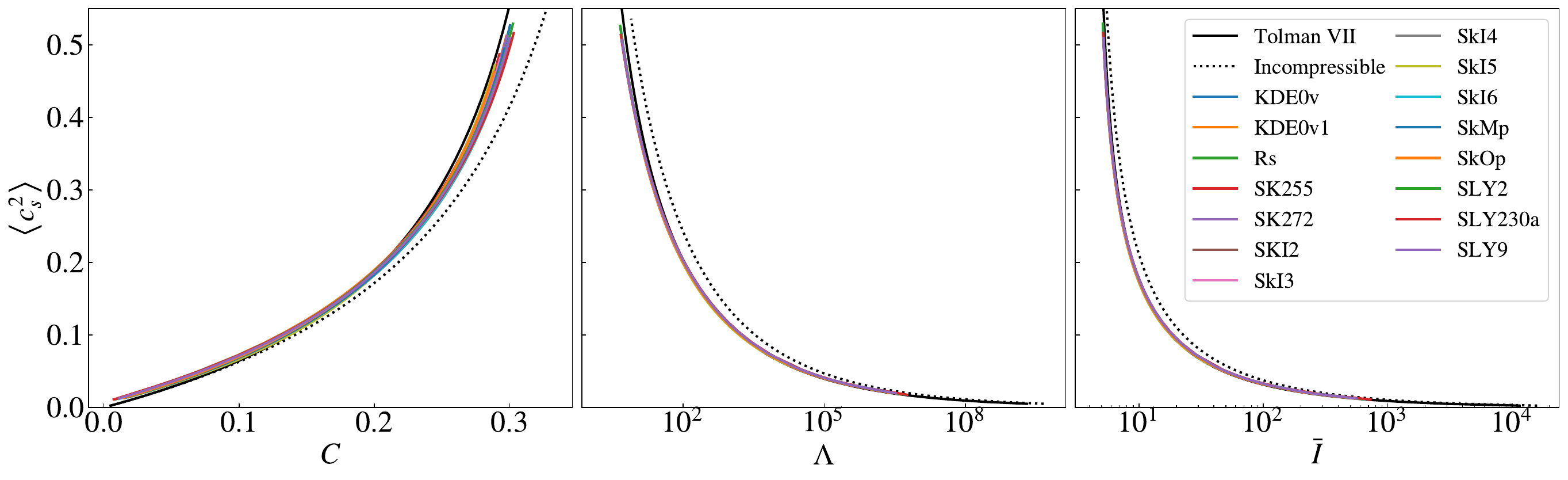}
   \caption{Relation between the average speed of sound squared (or, equivalently, the ratio of central pressure to central energy density, $\avgcs = p_c/\epsilon_c$) and the compactness ($C$), tidal deformability ($\Lambda$), and dimensionless moment of inertia ($\bar{I}$) of NSs obeying a set of realistic EoSs (KDE0v \cite{Gulminelli:2015csa,Danielewicz:2008cm,Agrawal:2003xb}, KDE0v1 \cite{Gulminelli:2015csa,Danielewicz:2008cm,Agrawal:2003xb}, Rs \cite{Gulminelli:2015csa,Danielewicz:2008cm,Friedrich:1986zza}, SK255 \cite{Gulminelli:2015csa,Danielewicz:2008cm,Agrawal:2003xb}, SK272 \cite{Gulminelli:2015csa,Danielewicz:2008cm,Agrawal:2003xb}, SKI2-6 \cite{Reinhard:1999ut}, SkMp \cite{Gulminelli:2015csa,Danielewicz:2008cm,Bennour:1989zz}, SkOp \cite{Reinhard:1999ut}, SLY2 \cite{Chabanat:1997qh}, SLY230a \cite{Gulminelli:2015csa,Danielewicz:2008cm,Chabanat:1997qh}, and SLY9 \cite{Gulminelli:2015csa,Danielewicz:2008cm,Chabanat:1997qh},
    with tables obtained from the CompOSE repository). Curves for an incompressible fluid and the Tolman VII solution are also shown.  }
   \label{fig:I-Love-C-cs2}
\end{figure*}

Approximate universal relations between macroscopic observables and hydrodynamic quantities are also known to exist. One example is the relation between the pressure at densities $(1$--$2)\,\rho_{sat}$ and the radius and tidal deformability of a NS with a mass of $(1$--$1.4)M_\odot$ \cite{Lattimer:2000nx, Lim:2018bkq}, which has been used to constrain the EoS space at these densities using observational data \cite{Lattimer:2014sga, Drischler:2021kxf}. In a recent study \cite{Saes:2021fzr}, a novel approximate universal relation of this type was established. This relation links the ratio between the central pressure and the central energy density, denoted as $p_c/\epsilon_c$, to the compactness $C$, dimensionless moment of inertia $\Bar{I}$, and dimensionless tidal deformability $\Lambda$. This relation was shown to hold for a large range of realistic and parameterized EoSs commonly used to describe NSs, and is illustrated in Fig.~\ref{fig:I-Love-C-cs2} for a set of tabulated EoSs. Through these relations, a measure of $C$, $\Lambda$ or $\bar{I}$ would therefore yield an inference of $p_c/\epsilon_c$.

The perhaps surprising insensitivity of this relation to the choice of EoS, despite it connecting hydrodynamic quantities ($p_c/\epsilon_c$) to macroscopic observables, raises questions about the underlying reasons for its existence. Previous research, e.g.~ \cite{Yagi:2013awa,Yagi:2014qua,Sham:2014kea,Chan:2015iou}, has been dedicated to identifying the causes of the emergence of EoS-insensitive relations in NSs, particularly regarding the I-Love-Q one. In these studies, several aspects of the approximate universal relation were analyzed, such as analytic approximations for the observables, the contributions of different segments of the NS interior and EoS to the approximate universality, and which assumptions are more relevant to the insensitivity, among others. Inspired by the strategies adopted in these works,
we here study the physical origins of the quasi-universality of the $p_c/\epsilon_c$ --- $C/\Lambda/\bar{I}$ relation presented in \cite{Saes:2021fzr}. 

We begin by establishing that the ratio $p_c/\epsilon_c$ can be identified with the averaged speed of sound squared inside a NS, $\avgcs\!$, where the average is understood to be taken across the range of energy densities present inside the star. This reinforces the interpretation of this quantity as a measure of the mean NS stiffness. Next, we study the approximately universal $\avgcs$ -- $C/\Lambda/\bar{I}$ relations analytically, both in the Newtonian limit and in post-Minkowskian expansions. In the Newtonian limit, we find that the same relation, $\avgcs = C/2$, applies for various EoSs (i.e.~an incompressible fluid, an $n=1$ polytrope, and a Tolman VII density profile \cite{Tolman:1939jz}), which, despite their simplicity, are relevant to the description of NSs. We interpolate between these models by considering two one-parameter families of EoSs, namely, a family of polytropes and a family of generalized Tolman VII profiles, the latter of which we introduce in this paper. We show that, in the Newtonian limit, the \avgcs -- $C$ relation is remarkably flat over the range of model parameters most relevant to the description of NSs, with an EoS-dependence of only $\sim 3\%$ and $\sim 6\%$ for polytropic EoSs and the generalized Tolman VII model, respectively (cf. Fig. \ref{fig:newtrelations} in Sec.~\ref{sec:newt_poly}). 

We confirm and extend the above conclusions by considering NSs in the post-Minkowskian expansion, expressing solutions as power series in the NS compactness. We find that the \avgcs-- $C$ relation becomes increasingly sensitive to the EoS as one includes higher powers of compactness (i.e.~higher order GR contributions). The EoS-dependence increases to $\sim 9\%$ for both the polytrope family and the generalized Tolman VII family, when considering up to third-order contributions in compactness for all stable NSs with these EoSs (cf. Figs. \ref{fig:PMpolyCcs} and \ref{fig:PMGTVIICcs} in Sec.~\ref{subsubsec:Poly}). We assess to which point these semi-analytic results for simple, one-parameter families of EoSs can be informative of realistic EoSs. We conclude this is true as long as the star is sufficiently compact ($C \gtrsim 0.1$).

With an understanding of the \avgcs-- $C$ relation under our belts, we consider the $I$-Love-\avgcs relation. We conduct a similar study in the Newtonian limit and through post-Minkowskian expansions. Unlike in  \avgcs-- $C$ case, we find that the $I$-Love-\avgcs relation presents the same degree of universality in the Newtonian limit (cf. Figs. \ref{fig:newtrelations_I} and \ref{fig:newtrelations_lambda} in Sec.~\ref{subsubsec:genT7}) and in the presence of increasing GR contributions (cf. Figs. \ref{fig:PMGTVIIIcs} and \ref{fig:PMGTVIIlambdacs} in Sec.~\ref{subsec:PMexp}). For example, we find that the sensitivity of the $I$-Love-\avgcs relation to variations of the generalized Tolman VII family remains at the $\sim 10-11\%$ level to all orders in compactness.

The remainder of this paper presents the details that support the conclusions summarized above and it is organized as follows. 
In Sec. \ref{sec:understandingalpha}, we elucidate the physical significance of $p_c/\epsilon_c$ as the average speed of sound squared. In Secs. \ref{sec:Ccs2} and \ref{sec:Ilovecs2}, we delve into the origins of the \avgcs -- $C$ relation and $I$-Love-\avgcs relations, correspondingly, following a post-Minkowksian approach. Section \ref{sec:conclusions} summarizes our conclusions. 

\section{$p_c/\epsilon_c$ as a measure of the average NS stiffness} 
\label{sec:understandingalpha}

Let us begin by providing a new interpretation for the ratio $p_c/\epsilon_c$ between the central pressure and central energy density.
As argued in ~\cite{Saes:2021fzr}, this ratio can be physically interpreted as a measure of the mean stiffness of nuclear matter inside a NS, indicative of the global growth of pressure with respect to the energy density.
Indeed, let us assume that the energy density is zero when the pressure is null, i.e., $\epsilon(p=0)=0$, so that then
\begin{align}
\label{eq:alphacs2cont}
\frac{p_c}{\epsilon_c} = \frac{1}{   \epsilon_c} \int_0^{\epsilon_c} \frac{dp}{d\epsilon} d\epsilon = \frac{1}{c^2
\epsilon_c} \int_0^{\epsilon_c} c_s^2 d\epsilon =:\avgcs\,.
\end{align}
The quantity \avgcs is the dimensionless speed of sound squared averaged over the range of energy densities present inside the NS. This expression reinforces the interpretation provided in \cite{Saes:2021fzr} since the speed of sound is a typical indicator of the stiffness of an EoS.

\begin{figure}[h]
\includegraphics[width=0.4\textwidth]{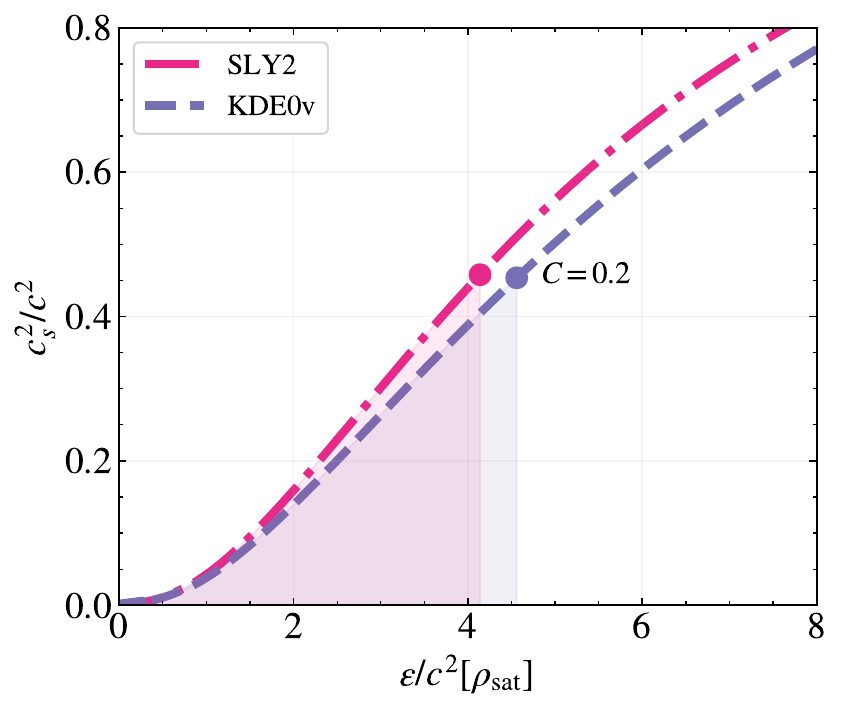}
   \caption{Speed of sound as a function of the energy density for two EoSs with varying stiffness, KDE0v \cite{Danielewicz:2008cm,Gulminelli:2015csa,Agrawal:2005ix} and SLY2 \cite{Danielewicz:2008cm,Gulminelli:2015csa,Chabanat:1997qh}
   The filled circles indicate the central energy density and the speed of sound squared that corresponds to a NS with compactness $ C = 0.2$ for each EoS. } 
   \label{fig:cs2_explanation}
\end{figure}

A visual representation of this average is shown in Fig. \ref{fig:cs2_explanation}, where we present the function $c_s^2(\epsilon)$ for two representative EoSs. The filled circles in this figure indicate the central energy density and the maximum speed of sound inside a star with $C=0.2$. The speed of sound squared, averaged over the energy density, is then nothing but the ``area under the curve'' (i.e.~the shaded regions) divided by the central energy density (because the energy density at the surface is zero).

The approximate universality between \avgcs and $C$ is then connected to the fact that two NSs with different EoSs but the same compactness will typically have different central energy densities. As we can see from Fig.~\ref{fig:cs2_explanation}, as one considers an EoS with a smaller stiffness, one naturally decreases the rate at which the speed of sound increases as a function of energy density in the NS interior. This, then, implies that, to obtain the same compactness as a NS with a stiffer EoS, the central energy density must be higher. This naturally increases the area under the $c_s^2$-$\epsilon$ curve, in such a way that the ratio of this new area to the new central energy density is roughly unchanged. To beyond these qualitative considerations, in the remainder of this paper, we analyze the $\avgcs$ -- $C/\Lambda/\bar{I}$ relations in a post-Minkowksian framework.

\section{\avgcs -- $C$  relation}
\label{sec:Ccs2}

To explore the underlying causes behind the quasi-universality of the $\avgcs$ -- $C$ relation, in this section, we first assess its presence within analytic solutions for the TOV equation \cite{Oppenheimer:1939ne, Tolman:1939jz}, namely, the incompressible, Tolman VII and Buchdahl solutions.
The common Newtonian limit of the $\avgcs$ -- $C$ relation in those cases motivates us to follow this up with a study of that relation for 2 one-parameter families of EoSs: polytropes and a generalized version of the Tolman VII EoS, first in the Newtonian limit and then within a post-Minkowskian framework.
Finally, we contrast these simple models with realistic EoSs and discuss in which regime the results obtained for the former can be informative of the latter.

\subsection{Exact Solutions in GR}
\label{subsec:ExactSolGR}

While several known analytic solutions to the TOV equations exist, not all of them yield relevant descriptions for NSs. Here, as in~\cite{Lattimer:2000nx}, we shall present the \avgcs -- $C$ relation for three exact solutions of the TOV equations: the incompressible fluid, a Tolman VII fluid, and a Buchdahl fluid. For concreteness, and to further establish notation, we review the TOV equations in GR in the appendix \ref{subap:GeneralRelativity}. Detailed formulae for the models we consider can be found in appendix \ref{subap:AnalyticalSolutions}.

\subsubsection{Incompressible fluid}

We first examine the incompressible case, which describes a fluid with constant density. The EoS is given by
\begin{align}
\label{eq:incompressibleeos}
\epsilon(p)  & = \epsilon_c = \text{constant},
\end{align}
which leads to the simple analytic relation\footnote{Although an incompressible fluid does not satisfy the condition $\epsilon(p=0)=0$ required for Eq.~(\ref{eq:alphacs2cont}) to hold, and has a diverging speed of sound, throughout the paper we adhere to the definition $\avgcs \equiv p_c/\epsilon_c$.} (cf. appendix \ref{subap:AnalyticalSolutions})
\begin{align}
\label{eq:meancs2incompressible}
\avgcs  = \frac{p_c}{\epsilon_c}  & =\frac{1 - \sqrt{1 - 2 C}}{3\sqrt{1-2C} - 1}.
\end{align}
This can be expanded for small values of compactness to find 
\begin{align}
\label{eq:inc_aprox}
\avgcs \approx\frac{C}{2}+C^2+\mathcal{O}(C^3).
\end{align}

Relation \eqref{eq:meancs2incompressible} presents some interesting, albeit well-known behavior. First, the averaged speed of sound diverges when $C \to 4/9$, which corresponds to the Buchdahl limit, i.e.~an equilibrium sequence of incompressible solutions to the TOV equations ceases to exist at this value of compactness. 
Second, the average speed of sound squared exceeds unity when $C > 3/8$. For extremely non-relativistic stars, the average speed of sound tends to zero, and, for weakly-gravitating stars, it scales with compactness. The radius of convergence of the post-Minkowskian expansion \eqref{eq:inc_aprox} is simply the Buchdahl limit, as one can confirm by inspection or through the ratio test. 

\subsubsection{Tolman VII fluid}

The Tolman VII solution, introduced in~\cite{Tolman:1939jz}, is characterized by the density profile
\begin{align}
\label{eq:TVIIdef}
      \epsilon(r)  & = \epsilon_c \left[1 - \left(\frac{r}{R}\right)^2\right].
\end{align}

Again, the averaged speed of sound squared can be computed analytically,
\begin{align}
\label{eq:meancs2tolman}
\avgcs= \frac{1}{15}  \left(\frac{2 \sqrt{3} \tan \phi_c}{\sqrt{C}}-5\right),
\end{align}
where $\phi_c = \phi_c(C)$ is given by Eq.~\eqref{eq:phi_c}.
This expression can be Taylor expanded in small $C$ to obtain
\begin{align}
\label{TVIIlim}
     \avgcs\approx \frac{C}{2}+\frac{133 C^2}{120}+\mathcal{O}(C^3).
\end{align}
From Eq.~\eqref{eq:meancs2tolman} we see that the central pressure, and consequently, the averaged squared speed of sound becomes infinite when the compactness reaches the point where $\phi_c(C)=\pi/2$, which corresponds to $C\approx0.386$. This value of the compactness is lower than the Buchdahl limit of incompressible fluids, and it defines the end of the equilibrium branch of the Tolman VII solution. However, before diverging, \avgcs reaches and subsequently exceeds unity at $C\approx0.335$, which sets the limit beyond which the fluid becomes superluminal, on average. The Taylor series \eqref{TVIIlim} for this analytic relation converges for $C \lesssim 0.386$, as one can check with the ratio test, which coincides with the value beyond which the central pressure diverges.

\subsubsection{Buchdahl fluid}

The Buchdahl EoS \cite{1967ApJ...147..310B} is given by
\begin{align}
\label{eq:buchdaldef}
\epsilon(p) = 12(p_* p)^{1/2} - 5p,
\end{align}
where $p_*$ is a constant. This equation of state reduces, in the Newtonian limit, to a $n=1$ polytrope, $p \propto \epsilon^2$.

As shown in appendix \ref{subap:AnalyticalSolutions}, the averaged speed of sound squared can be cast in this case as
\begin{align}
\label{eq:meancs2buchdal}
    \avgcs = \frac{C}{2-5 C},
\end{align}
and for small values of compactness:
\begin{align}
\label{eq:cs2expbuch}
    \left<c_s^2\right> \approx \frac{C}{2}+\frac{5 C^2}{4}+\mathcal{O}(C^3).
\end{align}
From Eq.~\eqref{eq:meancs2buchdal} one sees that \avgcs diverges when $C = 2/5$. Contrary to the previous cases, this is not due to the divergence of the central pressure [which is given by the regular expression~\eqref{eq:pcbuchdal}], but to the fact that the central energy density becomes null when $C = 2/5$. This will be the limiting factor that marks the end of the equilibrium sequence for the Buchdahl solution. 
Before that value, when $C=1/3$, the average squared speed of sound becomes greater than one, and the fluid becomes superluminal, on average. The expansion for low compactness values, Eq. \eqref{eq:cs2expbuch}, has a convergence radius of $C=2/5$, as expected from the analytic expression, and as can also be found via the ratio test. 

As a final remark, notice that, for the Buchdahl solution, the relation between \avgcs and $C$, given in Eq.~\eqref{eq:meancs2buchdal}, is exactly the ratio of two linear polynomials, which succinctly captures both the $\avgcs \to C/2$ limit as $C \to 0$ and the pole at $C=2/5$. This same structure, but suitably generalized, could provide useful fitting formulae for the case of realistic EoSs.

\subsubsection{Comparing solutions}

From the different \avgcs -- $C$ analytic relations found above, perhaps the most noteworthy feature is their common Newtonian limit.
The Taylor expansions of the analytic expressions shown above imply that $\avgcs = C/2 + {\cal{O}}(C^2)$, and the coefficient of the $C^2$ term is of order unity. This observation leads us to speculate that the emergence of universality in this limit may be attributed to the predominance of Newtonian effects over relativistic effects, at least for EoSs close to those considered here. To further explore this point, in the next subsection, we study the \avgcs -- $C$  relation in the Newtonian limit for two families of one-parameter EoSs, which interpolate between the models considered in this subsection. We will later perform a post-Minkowskian expansion to elucidate the role of GR in the quasi-universality. 


\subsection{One-Parameter Families of EoSs: \\ Newtonian Limit} 
\label{subsec:NewtRelation1}

\begin{figure}
\includegraphics[width=0.5\textwidth  ]{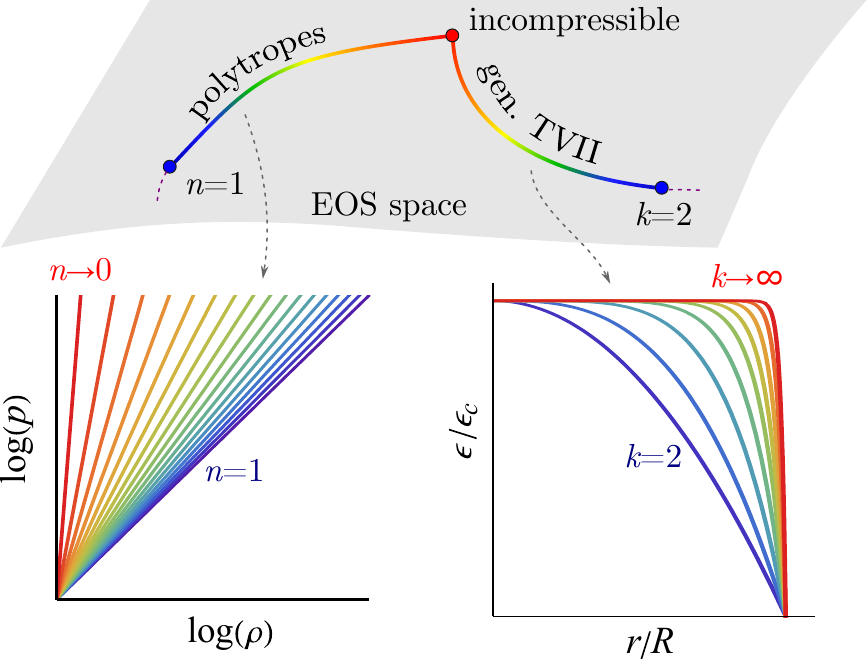}
   \caption{Illustration of the 2 one-parameter families to be considered along this work, namely, polytropes with $n\in[0,1]$ and a generalized Tolman VII model which interpolates between the original Tolman VII solution ($k=2$) and an incompressible density profile ($k\to \infty$).}
   \label{fig:EOSspace}
\end{figure}

Given the common Newtonian limit of the averaged speed of sound squared, regardless of the EoS considered above, it is of interest to study the Newtonian limit directly for other EoSs. 
By ``Newtonian limit,'' we here mean the leading-order expansion of the equations of structure when $c \to \infty$, or equivalently, when $G \to 0$ (see appendix \ref{subap:Newtonian}). Naturally, it is impractical to fully cover the space of EoS, as an infinite number of parameters would be needed. Instead, in what follows we present the \avgcs -- $C$  relation for 2 one-parameter families of EoS, namely, a polytropic EoS and a generalized version of the Tolman VII solution. As pictorially illustrated in Fig.~\ref{fig:EOSspace}, these families interpolate between (the Newtonian limit) of the models considered in Sec.~\ref{subsec:ExactSolGR}. By varying the EoS parameters that characterize them, one can get a handle on whether the universal Newtonian limit found for the three models in Sec.~\ref{subsec:ExactSolGR} is preserved or not across (certain paths in) the EoS space.

\subsubsection{Polytropic EoSs} \label{sec:newt_poly}

A polytropic EoS is defined as
\begin{align} 
\label{eq:polytrope}
    p = K \, \rho^{ 1 + \frac{1}{n}},
\end{align}
where $K$ is the polytropic constant, and $n$ is the polytropic index. For Newtonian polytropes, it is customary to define the dimensionless variables
\begin{align} \label{eq:ximu}
    \xi \equiv r/r_0, \qquad \mu^{(N,P)} \equiv m/m_0^{(N,P)},
\end{align}
where 
\begin{align} 
\label{eq:r0m0}
    r_0^2 \equiv \frac{(n+1) p_c}{4\pi G \rho_c^2 }, \qquad 
    m_0^{(N,P)} = 4 \pi \rho_c r_0^3,
\end{align}
are given in terms of the central rest mass density $\rho_c = \rho(r=0)$ and the superscript $(N,P)$ stands for ``Newtonian polytrope.'' The stellar radius is then given by $R = r_0 \xi_1$ and is determined by finding the value of the dimensionless variable $\xi$ at which $p(\xi=\xi_1)=0$. Similarly,  the stellar mass is given by $M = m_0 \mu_1^{(N,P)}$, where $\mu_1^{(N,P)} = \mu^{(N,P)}(\xi=\xi_1)$.

With this in hand, we can now compute the averaged speed of sound squared. From Eq.~(\ref{eq:r0m0}) it follows that 
\begin{align}
    \frac{G m_0}{r_0c^2} = (n+1) \frac{p_c}{\rho_c c^2} = (n+1) \avgcs,
\end{align}
so that the stellar compactness is linearly related to \avgcs via
\begin{align}
    \avgcs = C \left[(n+1) \frac{\mu^{(N,P)}_1}{\xi_1}\right]^{-1}.
\end{align}
In this expression, it is evident that both $(n+1)$ and $\mu_1^{(N,P)}/\xi_1$ are highly dependent of the EoS parameter $n$. Consequently, any universality must arise from the combination of both terms. 

\begin{figure}
\includegraphics[width=0.4\textwidth  ]{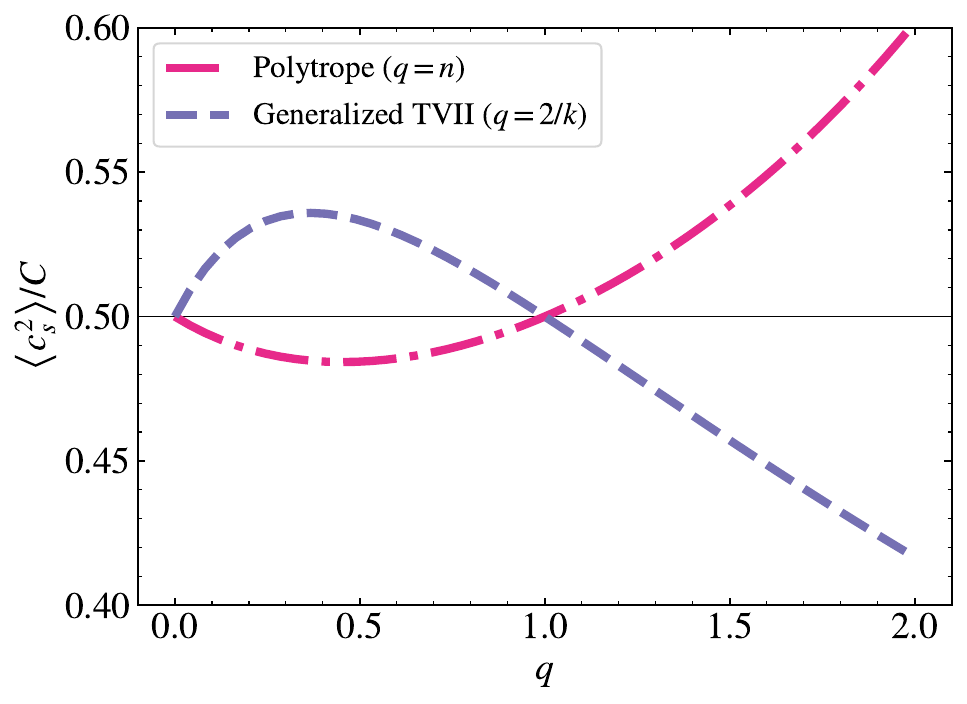}
   \caption{Ratio between the average speed of sound squared and the compactness for 2 one-parameter families
   of solutions to the Newtonian equations of structure, the polytropic EoS (in pink) and the generalized Tolman VII solution (in dashed cyan). We use the polytropic index $n$ to parameterize the former,
   and for the latter, we use the quantity $2/k$, which reduces to the incompressible case when $2/k = 0$ and to the Tolman VII case when $2/k = 1$.}
   \label{fig:newtrelations}
\end{figure}

The polytropic index values that suitably approximate the behavior of realistic EoSs for NSs are limited to $n \in [0,1]$. For the boundaries, $n=0$ (the incompressible limit) and $n=1$ (the Newtonian limit of a Buchdahl fluid), the equations of hydrostatic equilibrium can be analytically solved to find
\begin{align}
    n = 0 \longrightarrow & \quad p(\xi) = p_c \left( 1 - \frac{\xi^2}{6}\right), \quad \mu^{(N,P)}(\xi) = \frac{\xi^3}{3}, \nonumber \\
     &\quad \xi_1 = \sqrt{6}, \quad \mu_1^{(N,P)} = 2\sqrt{6}.
    \\
    n = 1 \longrightarrow & \quad p(\xi) = p_c \left( \frac{\sin \xi}{\xi} \right)^2, \quad \mu^{(N,P)}(\xi) = \sin\xi -\xi \cos\xi, \nonumber \\
    & \quad \xi_1 = \pi, \quad \mu_1^{(N,P)} = \pi.
\end{align}
Thus, one sees that in both cases
\begin{align}
\label{eq:cs2relationpoly}
   \left<c_s^2\right> = \frac{C}{2} \qquad (n = 0 \; {\textrm{or}} \;  1).
\end{align}
In fact, by solving the Lane-Emden equation (i.e.~the Newtonian version of the TOV equation in the reduced variables defined above) numerically for values of $n \in [0,1]$, we find that the maximum fractional difference in $\avgcs\!/C$ with respect to a fiducial EoS with $n=0.5$ (the EoS median) is $\sim 3\%$.
This can be seen in  Fig. \ref{fig:newtrelations}, where we show the ratio $\avgcs\!/C$ for the extended interval $n \in [0,2]$. This ratio is notably flat for $n\in [0,1]$, with a stationary point at $n \approx 0.454$, but rapidly increases for higher values of $n$, eventually diverging as $n \to 5$. It thus becomes clear that the universality of the \avgcs -- $C$ relation in the Newtonian limit is restricted to relatively stiff EoS (i.e., to relatively low values of $n$).

\subsubsection{Generalized Tolman VII EoS}
\label{sec:genTVIIN}
Let us now consider fixing the density profile through a generalized Tolman VII EoS, which we define as
\begin{align}\label{eq:genTVIIN}
    \rho (r) = \rho_c \left[1 - \left(\frac{r}{R}\right)^k \right]\,.
\end{align}
The constant $k$ must here satisfy $k>1$ to ensure that $d\rho/dr|_{r=0} = 0$. When $k=2$, the above profile reduces to the Newtonian version of the Tolman VII profile shown in Eq.~\eqref{eq:TVIIdef}, which is thought to be a reliable approximation for the density profile inside NSs. As $k \to \infty$, we recover the flat profile characteristic of an incompressible fluid (see Eq.~\eqref{eq:incompressibleeos}). Therefore, Eq.~(\ref{eq:genTVIIN}) provides an interesting way of ``interpolating'' between the Tolman VII and the incompressible fluid profiles, which allows for the analytic analysis to be developed further. 

Let us define the dimensionless variables
\begin{align} \label{eq:xmu}
    x \equiv r/R, \qquad \mu^{(N,T)} \equiv \frac{m}{m_0^{(N,T)}}, 
\end{align}
where $R$ is the stellar radius, and the superscript $(N,T)$ indicates the Newtonian generalized Tolman VII solution, where we set 
\begin{align} \label{eq:m0N}
m_0^{(N,T)} \equiv 4\pi \rho_c R^3.
\end{align}
Then, for arbitrary values of $k$, the equation of hydrostatic equilibrium and the equation for the enclosed mass can be solved analytically to find
\begin{equation}
    p(x) = p_c \left[1 - \frac{\beta}{\avgcs} \left( \frac{x^2}{6} - \frac{x^{k+2} (k+6)}{3(k+2)(k+3)} + \frac{x^{2k+2}}{2(k+1)(k+3)} \right) \right], 
\end{equation}
\begin{equation}
    \mu^{(N,T)}(x) = \frac{x^3}{3} - \frac{x^{k+3}}{k+3},
\end{equation}
where anticipating the notation we will use in our post-Minkowskian calculations, we have defined $\beta\equiv G m_0/Rc^2$. 

With this in hand, we can now compute the average speed of sound squared. From the fact that $p(x=1) = 0$, and considering that
\begin{align}
\label{eq:betaCT7N}
    C = \frac{G M}{Rc^2} = \beta \mu^{(N,T)}(1) = \beta \left(\frac{1}{3} - \frac{1}{k+3}\right),
\end{align}
one finds the simple relation
\begin{align} \label{eq:CalphacT7N}
  \left<c_s^2\right> = \left(2 + \frac{1}{k} - \frac{3}{k+4}\right)^ {-1}C,
\end{align}
from where we see that
\begin{align}
      \left<c_s^2\right> = \frac{C}{2}\quad \textrm{for } k=2 \textrm{ and } k\to \infty.
\end{align}
For $k\in[2,\infty)$ in Eq.~(\ref{eq:CalphacT7N}), the maximum fractional difference in $\avgcs\!/C$ with respect to the EoS median is $\sim 6\%$, which characterizes the dependence of the relation with respect to $k$ in such interval. Here and below, when referring to the generalized Tolman VII model, the median is defined using a uniform probability distribution in the variable $q=2/k \in [0,1]$, i.e., it refers to a fiducial model with $q=0.5$ ($k=4$). The ratio $\avgcs\!/C$ is shown in Fig. \ref{fig:newtrelations} for the full range $k\in(1,\infty)$. Similarly to the polytropic case,  $\avgcs\!/C$ is relatively flat over the range $k\in[2,\infty)$, displaying a stationary point between the incompressible and the Tolman VII solutions, when $k = 2(1+\sqrt{3})\approx 5.46$, or equivalently, $q \approx 0.37$.

\subsection{One-Parameter Families of EoSs: \\ Post-Minkowskian Expansion}

In the previous subsections, we observed that the emergence of the quasi-universality of the \avgcs -- $C$ relation is rooted in the Newtonian limit and restricted to relatively stiff EoS. Furthermore, upon examining the exact solutions in GR, we noted that the contribution of higher-order terms becomes increasingly dependent on the EoS. To more precisely quantify the impact of GR on this dependence, we perform a post-Minkowskian expansion for the aforementioned one-parameter families of EoSs, which allows us to systematically study and analyze such effects semi-analytically.

\subsubsection{Polytropic EoSs} \label{subsubsec:Poly}

The relativistic generalization of the polytropic family of EoS is
\begin{align}
\label{eq:relpolytrope}
p(\rho) = K\rho^{1+\frac{1}{n}}, \quad \text{with } \quad \epsilon=\rho c^2 +np.
\end{align}
We will once more use the dimensionless quantities defined in Eq. \eqref{eq:ximu}, $\xi$ and $\mu^{(P)}$ here, where the superscript $(P)$ indicates that we are referring to the polytropic EoS. Additionally, we use the definitions $\bar{p} \equiv p/p_c$, $\bar{\rho} \equiv \rho/\rho_c$, $\bar{\epsilon} \equiv \epsilon/(\rho_c c^2)$, where the subscript $c$ stands for the central value of a particular variable. The relativistic structure equations can then be written as
\begin{align}
    \frac{d\mu^{(P)}}{d\xi} = \xi^2 \bar{\epsilon}, \qquad
    \frac{d \bar{p}}{d\xi} = - (n+1) \frac{(\bar{\epsilon} + \sigma \bar{p})(\mu^{(P)} + \sigma \xi^3 \bar{p})}{\xi^2 (1-2(n+1) \sigma \mu^{(P)}/\xi)}.
\end{align} 
The dimensionless parameter $\sigma \equiv p_c/(\rho_c c^2)$, is related to the average speed of sound squared via 
\begin{equation} \label{eq:cs2-sigma}
    \avgcs = \frac{p_c}{\rho_c c^2 + n p_c} = \frac{\sigma}{1+n\sigma}.
\end{equation}
In the Newtonian limit, we have that $\sigma \ll 1$, and therefore $\avgcs \sim \sigma \ll 1$ which then implies that all three quantities ($\avgcs$, $\sigma$, and $C$) are natural expansion parameters in the post-Minkowskian approximation. 

By considering $\sigma$ as the small parameter, we can search for power series solutions of the form
\begin{equation}
    \mu^{(P)}(\xi; \sigma) = \sum_{i=0}^N \mu_{(i)}^{(P)}(\xi) \sigma^i + \mathcal{O}(\sigma^{N+ 1}),
\end{equation}
\begin{equation}
    \bar{p}(\xi; \sigma) = \sum_{i=0}^N p_{(i)}(\xi) \sigma^i + \mathcal{O}(\sigma^{N+ 1}).
\end{equation}
Additionally, the dimensionless radius of the star and dimensionless mass of the star can also be expanded in the form
\begin{equation} \label{eq:xi1_expand}
\xi_1(\sigma) = \sum_{i = 0}^N \xi_{1,i} \sigma^i, \qquad   \mu_1^{(P)}(\sigma) = \sum_{i = 0}^N \mu_{1,i}^{(P)} \sigma^i,  
\end{equation}
and can be obtained by solving the resulting equations perturbatively (see Appendix \ref{ap:PMexpansions} for details). 

The averaged speed of sound squared as a function of the compactness can now be found as follows. The compactness is obtained by evaluating
\begin{equation}
    C = (n+1) \sigma \frac{\mu_1^{(P)}(\sigma)}{\xi_1(\sigma)}
\end{equation}
as a power series in $\sigma$. This series can be rewritten in terms of \avgcs$\!$, and can subsequently be inverted into a power series in $C$. For example, for an $n=1$ polytrope, the next-to-leading-order term in the \avgcs -- $C$ relation can be obtained analytically, and the subsequent terms can be obtained numerically, namely
\begin{align}
    \avgcs \approx & \frac{C}{2} + \frac{C^2}{16} \left[ 8 + 9 \text{Ci}(\pi) - 9 \text{Ci}(3\pi) + 9 \ln 3 \right] + 2.97 C^3 +\mathcal{O}(C^4) \nonumber \\
    \approx & \frac{C}{2} + 1.15 C^2 + 2.97 C^3 +\mathcal{O}(C^4).
\end{align}
Observe that, similarly to the analytic solutions, the coefficient of the $C^2$ term is of order unity.

\begin{center}
\begin{figure}
\includegraphics[width=0.5\textwidth ]{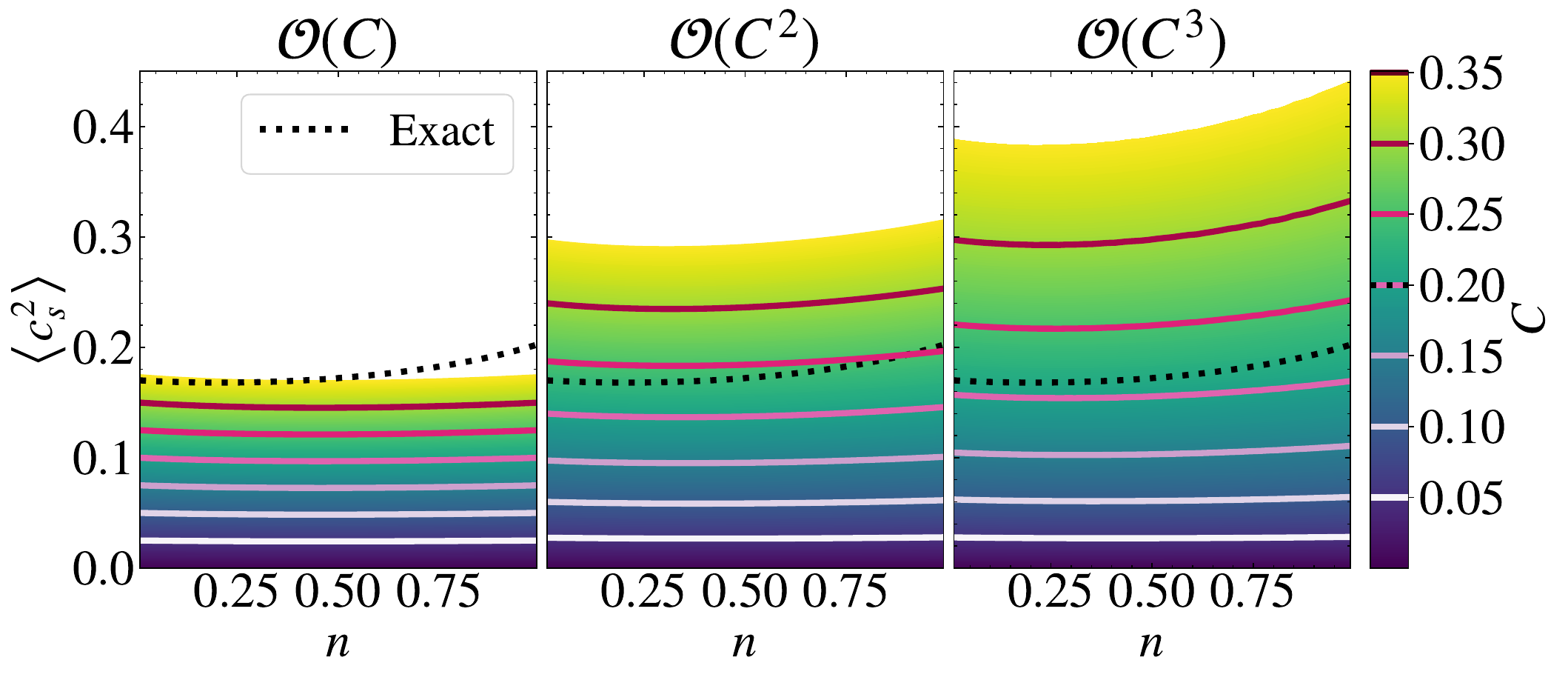}
   \caption{The variation of \avgcs in relation to the post-Minkowskian expansion for the polytropic equation of state, considering values of $n \in [0,1]$ along the bottom axis and representing different compactness levels with distinct colors. The expansion analysis starts from the first-order term and progressively incorporates higher-order contributions, up to the third order, proceeding from left to right. The black dotted line indicates the numerical solution in full GR for a NS with $C = 0.2$ for varying values of $n$.}
   \label{fig:PMpolyCcs}
\end{figure}
\end{center}

The post-Minkowskian expansion of the averaged speed of sound with $n\in[0,1]$ is shown in Fig. \ref{fig:PMpolyCcs}, which demonstrates the effects of higher-order terms on the \avgcs -- $C$ relation. A significant observation derived from this plot is that incorporating the $C^2$ and $C^3$ terms results in an increased sensitivity to the EoS in the $\avgcs$ -- $C$ relation. For a fiducial compactness of $C=0.2$, the maximal fractional deviation with respect to the EoS median, for polytropes with $n \in [0,1]$, raises from $\sim 3\%$ at $\mathcal{O}(C)$ (Newtonian limit) to $\sim 7\%$ at $\mathcal{O}(C^2)$, $\sim 9\%$ at $\mathcal{O}(C^3)$, and $\sim 17\%$ in full GR. This observation is consistent with the notion that the insensitivity to the EoS (for $n \in [0,1]$), which stems from the Newtonian limit, deteriorates (i.e.~the \avgcs -- $C$ relation becomes more sensitive to the EoS) as GR contributions are included.

\subsubsection{Generalized Tolman VII}

Let us now consider the case of the generalized Tolman VII density profile, defined in the relativistic regime as
\begin{equation} \label{eq:genTVIIR}
    \epsilon(r) = \epsilon_c \left[ 1 - \left( \frac{r}{R} \right)^k \right],
\end{equation}
where $k>1$. This family of solutions interpolates between the Tolman VII solution, for $k=2$, and the incompressible case, that once again corresponds to $k \to \infty$. Taking the Newtonian limit, $\epsilon = \rho c^2$, one recovers the results presented in Sec.~\ref{sec:genTVIIN}. As done before, we define the dimensionless quantities in Eq.~(\ref{eq:xmu}), but replacing Eq.~(\ref{eq:m0N}) by 
\begin{equation}
    m_0^{(T)} \equiv 4\pi \epsilon_c R^3.
\end{equation}
We use the same definition of $\beta$, with $\bar{p} = p/p_c$ and $\bar{\epsilon} = \epsilon/\epsilon_c$. The equation for the enclosed mass can be directly integrated and results in 
\begin{equation} \label{eq:muTVII}
    \mu^{(T)}(x) = \frac{x^3}{3} - \frac{x^{k+3}}{k+3},
\end{equation}
where the superscript ${(T)}$ indicates the generalized Tolman VII solution and where once again the compactness and the parameter $\beta$ are linearly related by
\begin{equation}
    C = \frac{G M}{Rc^2} = \beta \mu^{(T)}(1) = \beta \left( \frac{1}{3} - \frac{1}{k+3} \right).
\end{equation}
The dimensionless pressure then obeys the differential equation
\begin{equation}
    \frac{d \bar{p}}{dx} = - \frac{\beta}{\avgcs} \frac{(\bar{\epsilon}+\avgcs \bar{p})(\mu^{(T)} + \avgcs x^3 \bar{p})}{x^2(1-2\beta \mu^{(T)}/x)}.
\end{equation}

Let us now choose stellar compactness as our small post-Minkowskian expansion parameter to obtain power series approximations to the solution for the pressure. 
\begin{equation} \label{eq:psexpansionC}
    \bar{p}(x;C) = \sum_{i=0}^N p_i(x) C^i, \qquad
    \avgcs(C) = \sum_{i = 0}^N d_{i+1} C^{i+1},
\end{equation}
 where we write the relevant boundary condition as $\bar{p}(0;C) = 1$, while the requirement that $\bar{p}(1;C) = 0$ determines the coefficients in the expansion of $\avgcs(C)$. For the generalized Tolman VII models, the structure equations can be solved analytically, order by order, for any $k$ and up to high values of $N$. Here we show such analytic solution up to order $C^3$ for generic values of $k$
\begin{align}
\label{eq:PMGTVIICcs}
    \avgcs &= \frac{1+ 2 \left(\frac{2}{k}\right)}{\left( 1+ \left( \frac{2}{k}\right)\right)\left( 2+ \left( \frac{2}{k}\right)\right)} C
    \nonumber \\
    &+ \frac{ 1+\frac{17}{3}\left(\frac{2}{k}\right)+\frac{239}{24}\left(\frac{2}{k}\right)^2+\frac{133 }{24}\left(\frac{2}{k}\right)^3}{3 \left( 1+ \left( \frac{2}{k}\right)\right)^2 \left(3+8\left(\frac{2}{k}\right)+4\left(\frac{2}{k}\right)^2\right)}C^2 
    \nonumber \\
    &+ \frac{17}{2}
    \frac{1+\frac{33}{4}\left(\frac{2}{k}\right)+\frac{3165}{136}\left(\frac{2}{k}\right)^2+\frac{458}{17}\left(\frac{2}{k}\right)^3+\frac{1497}{136}\left(\frac{2}{k}\right)^4}{ \left(1+\left(\frac{2}{k}\right)\right)^3 \left(4+15 \left(\frac{2}{k}\right)+9\left(\frac{2}{k}\right)^2\right)} C^3 
    \nonumber \\
    &+ \mathcal{O}(C^4).
\end{align}

\begin{center}
\begin{figure}
\includegraphics[width=0.5\textwidth ]{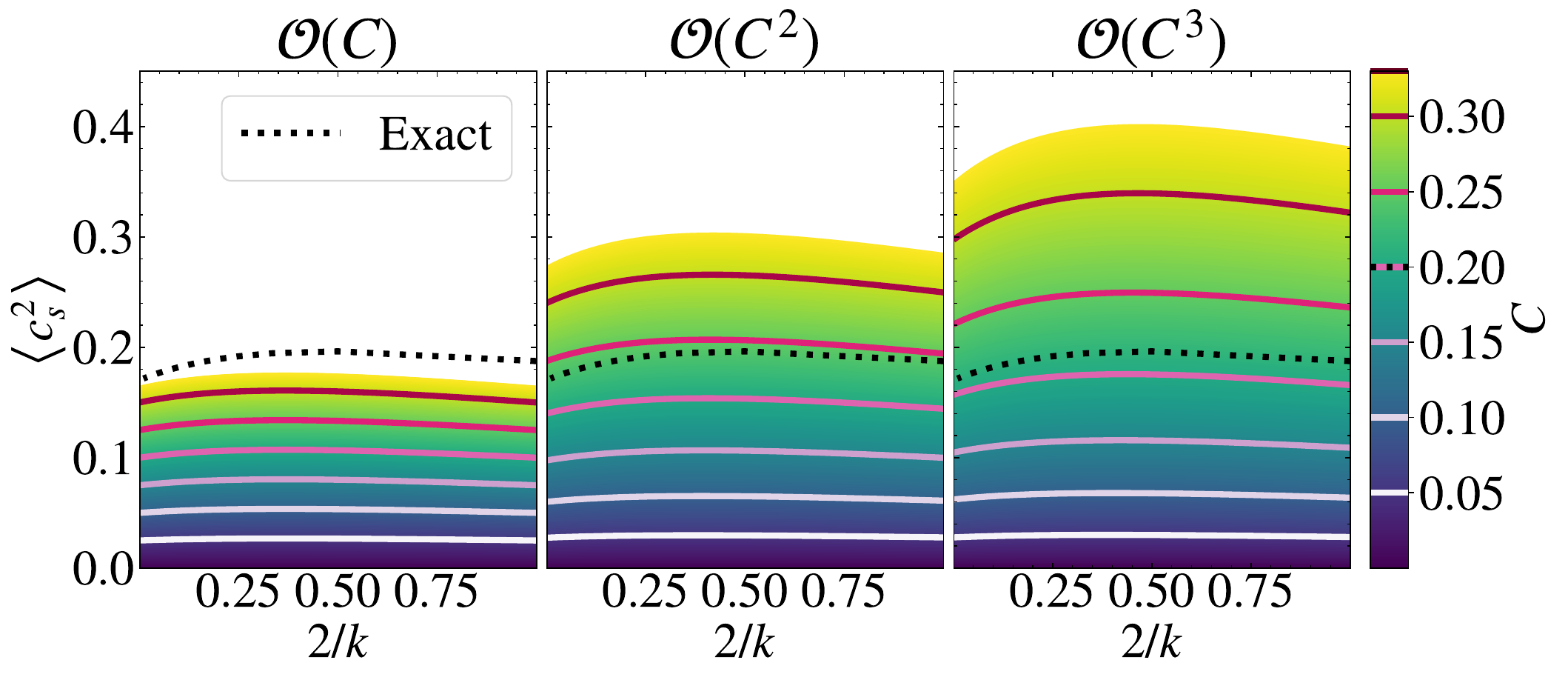}
   \caption{The variation of \avgcs in relation to the post-Minkowskian expansion for the generalized Tolman VII solution, considering values of $k \in [2,\infty)$ along the bottom axis and representing different compactness levels with distinct colors. The expansion analysis starts from the first-order term and progressively incorporates higher-order contributions, up to the third order, proceeding from left to right. The black dotted line indicates the numerical solution in full GR for a NS with $C = 0.2$ for varying values of $k$.}
   \label{fig:PMGTVIICcs}
\end{figure}
\end{center}

The influence of higher-order post-Minkowskian terms (i.e.~higher order in $C$ terms) on the EoS-dependence of the  $\avgcs\! -C$ relation is shown in Fig.~\ref{fig:PMGTVIICcs}. 
Introducing the $C^2$ and $C^3$ terms leads to an increased variability in the $\avgcs\! -C$ relation. This observation again, as in the polytropic case, aligns with the notion that the insensitivity to the EoS (for generalized Tolman VII models), which originates in the Newtonian limit, diminishes as GR contributions are added.
For a fiducial compactness of $C = 0.2$, the maximum fractional deviation of \avgcs with respect to the EoS median, for $q = 2/k \in [0,1]$, increases from $\sim 6\%$ at $\mathcal{O}(C)$ (Newtonian limit) to $\sim 8\%$ at $\mathcal{O}(C^2)$, $\sim 9\%$ at $\mathcal{O}(C^3)$, and $\sim 13\%$ in full GR.

\subsection{Comparison to Realistic EoSs}
 
To conclude this section, let us assess how useful the results obtained above (for simple, one-parameter families of EoSs) can be for understanding the $\avgcs\! -C$ relation when using realistic EoSs. To address this question, we consider a set of 15 tabulated EoSs, which is the same as that employed in Fig.~\ref{fig:I-Love-C-cs2}. In particular, we consider the KDE0v, KDE0v1, Rs, SK255, SK272, SKI2-6, SkMp, SkOp, SLY2, SLY230a, and SLY9 EoSs, with tables obtained from the CompOSE repository.

Realistic EoSs encode the rich behavior of nuclear matter from the outer crust to the inner core. The adiabatic index, $\Gamma = d (\log p)/ d (\log \rho)$ predicted by these EoSs typically spans a large interval, starting from $\sim 1.33$ at the outer crust, decreasing after neutron drip, and then increasing considerably (typically above 3) after the crust-core interface \cite{Douchin:2001sv}. 
Naturally, a single polytrope is not capable of capturing the full behavior of the nuclear EoS, nor is the density profile fully described by a simple analytic function as in Eq.~(\ref{eq:genTVIIR}).
Still, it will be useful to construct effective polytropic and generalized Tolman VII models to NSs described by our set of realistic EoSs, in order to verify to which extent (or in which domain) our results for the former models can be used to build intuition for the latter.

From the many possible definitions of an effective polytropic index of an EoS inside a NS, we select the simplest one, consisting of 
\begin{equation}
    n_\text{eff} = \frac{1}{\Gamma(\rho_c) – 1}\,,
\end{equation}
and thus, $n_\text{eff}$ is fully determined by the adiabatic index at the center of the star. Additionally, we define an effective generalized-Tolman-VII exponent, $k_\text{eff}$, by fitting an expression of the form of Eq.~(\ref{eq:genTVIIR}) to the energy-density profile of a given NS. To assess the adequacy of these effective descriptions, we introduce the following error measure,
\begin{equation} \label{eq:error_norm}
E = \sqrt{ \left( \frac{\Delta \avgcs}{\avgcs} \right)^2 + \left( \frac{\Delta C}{C} \right)^2, }
\end{equation}
where $\Delta$ denotes the difference between the quantities predicted by the effective model (the polytropic or generalized Tolman VII mimickers) and the original tabulated EoSs, which is then divided by the quantity predicted by the original EoS.

\begin{figure}
\includegraphics[width=0.45\textwidth]{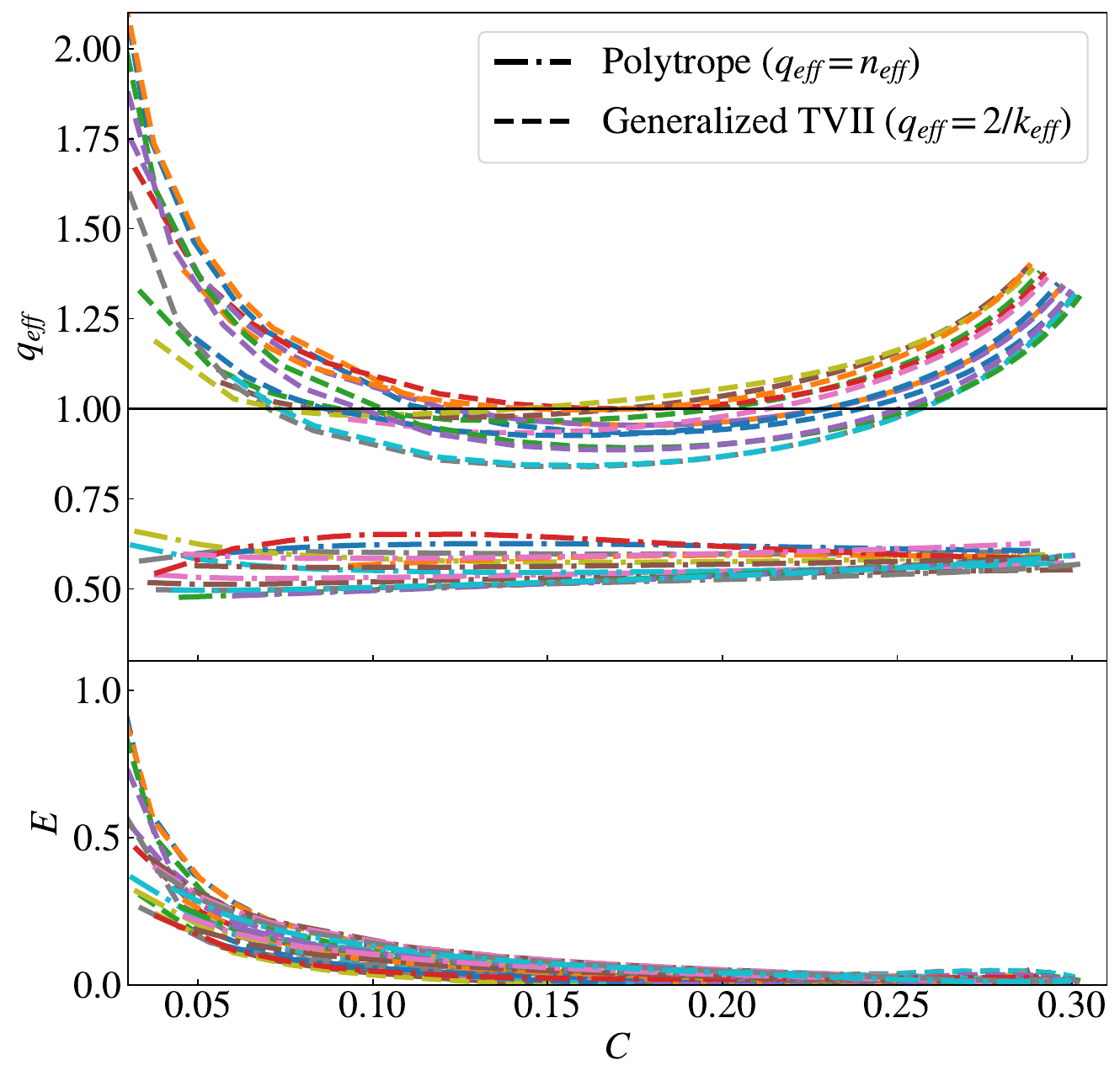}
   \caption{Effective indices and error norm as a function of the stellar compactness. The top panel shows the effective polytropic index $q_\text{eff} = n_\text{eff}$ (solid lines) and the effective generalized Tolman VII exponent $q_\text{eff} = k_\text{eff}$ (dashed lines) for fits to NSs described by a set of 15 tabulated EoSs. 
   The bottom panel shows the error measure, as defined in Eq.~(\ref{eq:error_norm}). }
   \label{fig:effective}
\end{figure}

\begin{figure}
\includegraphics[width=0.48\textwidth]{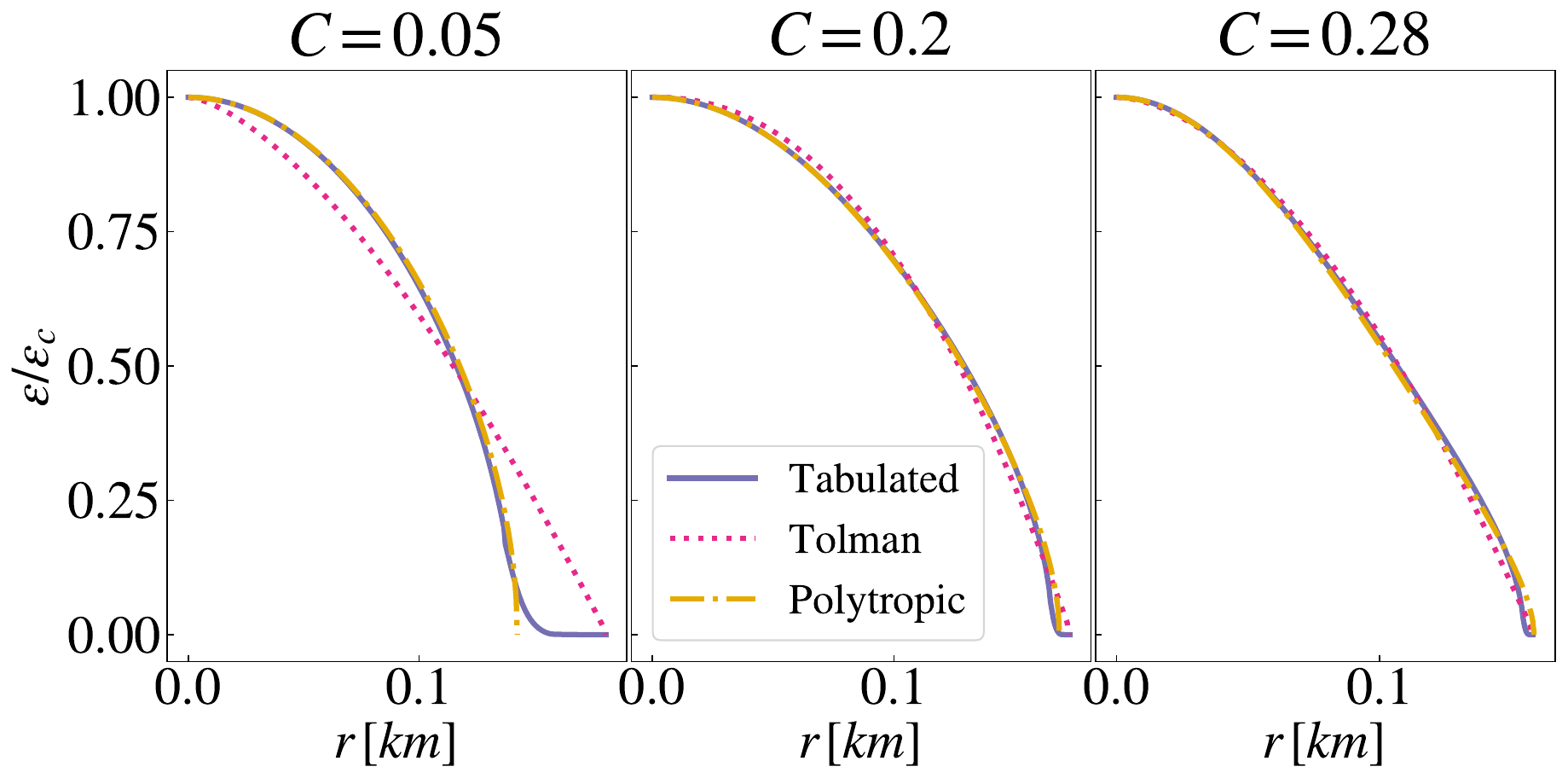}
   \caption{Energy density profiles for a low (left panel), medium (middle panel), and a high-compactness (right panel) NS. In each panel, the radial profile is shown for the tabulated (SLY9) EoS, the effective polytrope, and the effective generalized Tolman VII model, as defined in the main text.}
   \label{fig:density_profiles}
\end{figure}

Figure \ref{fig:effective} shows the effective indices $q_\text{eff} = \{n_\text{eff}, 2/k_\text{eff}\}$ as a function of the stellar compactness, for our set of tabulated EoSs, while the bottom panel shows the error $E$ as defined in Eq.~(\ref{eq:error_norm}). In the range of compactness considered here, $n_\text{eff} \in (0,1)$ for a polytropic description of the EoS, but it fails to reasonably describe the NS properties for $C \lesssim 0.1$, as the error $E$ rapidly increases. Physically, this is due to the fact that the effective polytrope underestimates the stellar radius for low compactness NSs, for which the crust occupies a larger portion of the stellar volume. This can be visualized in Fig.~\ref{fig:density_profiles}, where the energy density profile is represented for very low ($C=0.05$), medium ($C=0.2$), and high-compactness ($C=0.28$) NSs. 

Analogously, an effective generalized Tolman VII model also fails to describe low compactness stars, with $k_\text{eff}$ eventually dropping below 1, which is physically unacceptable. By construction, these models preserve the radius of the star but may fail to reasonably predict its mass (cf. Fig.~\ref{fig:density_profiles}), giving rise to large errors for small values of $C$. 
For $C \gtrsim 0.1$, $E \lesssim 0.1$ and our results for these simple one-parameter families of EoSs can be extended to interpret the case of realistic EoSs.

\section{I -- Love -- \avgcs}
\label{sec:Ilovecs2}

Alongside the near EoS-insensitive relation between \avgcs and $C$, \cite{Saes:2021fzr} presented a similar relation between the average speed of sound squared and both the dimensionless moment of inertia $\bar{I}$ and the dimensionless tidal deformability $\Lambda$, which is insensitive to the EoS at the $\sim 10\%$ level for realistic EoSs. Having explored the origins of the quasi-universality of the \avgcs -- $C$ relation, and arguing that, for relevant models that describe the NS interior, this quasi-universality emerges from the Newtonian limit and subsequently decays with increasing contributions of GR, we will now investigate the $\avgcs$ -- $\bar{I}/\Lambda$ relations. Due to the lack of analytic solutions in GR for both these quantities, we start by analyzing the relations in the Newtonian regime and proceed to study their corresponding post-Minkowskian corrections.

Before we begin, let us define the dimensionless quantities that shall be used throughout this section. The dimensionless moment of inertia and the dimensionless tidal deformability are respectively defined as
\begin{align}
\label{eq:dimensionless}
    \bar{I} \equiv \frac{c^4 I}{G^2 M^3}, \qquad \Lambda= \frac{2 k_2}{3 C^5},
\end{align}
where $I$ is the moment of inertia and $k_2$ is the $l=2$ tidal Love number.

\subsection{Newtonian Limit} 

To delve into the question of the role of the Newtonian limit in the $\avgcs$ -- $\bar{I}/\Lambda$ relations, we study them in this limit and compare any universality, to that present in the \avgcs -- $C$ relation, or lack thereof. 

The moment of inertia in the Newtonian regime, which we shall name $I^{(N)}$, is defined as
\begin{equation}
\label{eq:Newt_I}
    I^{(N)} = \frac{8\pi}{3} \int_0^R  r^4 \rho(r) dr.
\end{equation}
The tidal deformability is found through the perturbation of a non-rotating equilibrium state. If a static quadrupolar tidal field perturbs such state, the Eulerian change $h_2(r)$ to the gravitational potential can be found through the Newtonian limit of Eq. (15) in \cite{Hinderer:2007mb}:
\begin{equation}\label{eq:HeqN}
    \frac{d^2 h_2}{dr^2} + \frac{2}{r} \frac{dh_2}{dr} - \left[ \frac{6}{r^2} - 4\pi G \rho \frac{d\rho}{dp} \right] h_2 = 0,
\end{equation}
and noticing that regularity at $r=0$ implies that $h_2(r) \sim r^2$ as $r \to 0$. Defining $y \equiv R h_2'(R)/h_2(R)$, the  Newtonian $l=2$ Love number can be computed from
\begin{equation}
\label{eq:Newtk_2}
    k_2^{(N)} = \frac{1}{2} \left( \frac{2-y}{y+3} \right),
\end{equation}
which is then related to the dimensionless tidal deformability via Eq. \eqref{eq:dimensionless}.
\subsubsection{Polytropic EoSs}

Let us first return to polytropic EoSs, defined in the Newtonian limit in Eq. \eqref{eq:polytrope}. For the analytic solutions presented for polytropes in Sec. \ref{subsec:NewtRelation1}, when $n = 0$ and $n = 1$, we obtain analytic solutions for the moment of inertia:
\begin{align}
     n = 0 \longrightarrow & \quad     \bar{I}^{(N,P)} =  \frac{2}{5}\frac{1}{C^2},\\
      n = 1 \longrightarrow & \quad  \bar{I}^{(N,P)} = \left(\frac{2}{3} - \frac{4}{\pi^2}\right) \frac{1}{C^2} \,,
\end{align}
which can then be related to the averaged speed of sound squared via 
\begin{align}
    n = 0 \longrightarrow & \quad   \avgcs  = \frac{1}{\sqrt{10}} \left(\frac{1}{ \bar{I}^{(N,P) }}\right)^{1/2}   ,\\
      n = 1 \longrightarrow & \quad \avgcs  \approx 0.26 \left(\frac{1}{ \bar{I}^{(N,P) }}\right)^{1/2} .
\end{align}
Unlike the case of the \avgcs-- $C$ relation, there is no common limit of the \avgcs-- $\bar{I}$ relation between the incompressible model and the $n=1$ polytrope. For polytropes with $n \in [0,1]$, the quantity $ \left(\bar{I}^{(N,P)}\right)^{1/2}\avgcs$ displays a fractional difference with respect to the EoS median of (at most) $\sim 14\%$. This quantity is shown in Fig. \ref{fig:newtrelations_I} as a function of $n \in [0,1]$.

Such EoS dependence is also present in the relation between the dimensionless tidal deformability and the average speed of sound squared. When $n=0$ and $n = 1$ we find that
\begin{align}
       n = 0 \longrightarrow & \quad   \Lambda^{(N,P)} =  \frac{1}{2 C^5}\,, \\
           n = 1 \longrightarrow & \quad   \Lambda^{(N,P)} = \frac{15-\pi^2}{3\pi^2}  \frac{1}{C^5}\,,
\end{align}
and thus
\begin{align}
       n = 0 \longrightarrow & \quad \avgcs \approx 0.44 \left(\Lambda^{(N,P)} \right)^{-1/5} \\
           n = 1 \longrightarrow & \quad  \avgcs \approx 0.35 \left(\Lambda^{(N,P)} \right)^{-1/5}.
\end{align}
For polytropes with $n \in [0,1]$, the quantity $ \avgcs \left( \Lambda^{(N,T)} \right)^{1/5}$ displays a fractional difference with respect to the EoS median of (at most) $\sim 14\%$. This quantity is shown in Fig. \ref{fig:newtrelations_I} as a function of $n \in [0,1]$.

\subsubsection{Generalized Tolman VII}\label{subsubsec:genT7}

Let us now examine the $\avgcs$ -- $\bar{I}/\Lambda$ relations using the generalized Tolman VII profile. The relevant EoS expression and equilibrium equations pertaining to this solution were discussed in Sec. \ref{sec:genTVIIN}.

\begin{figure}
\includegraphics[width=0.4\textwidth  ]{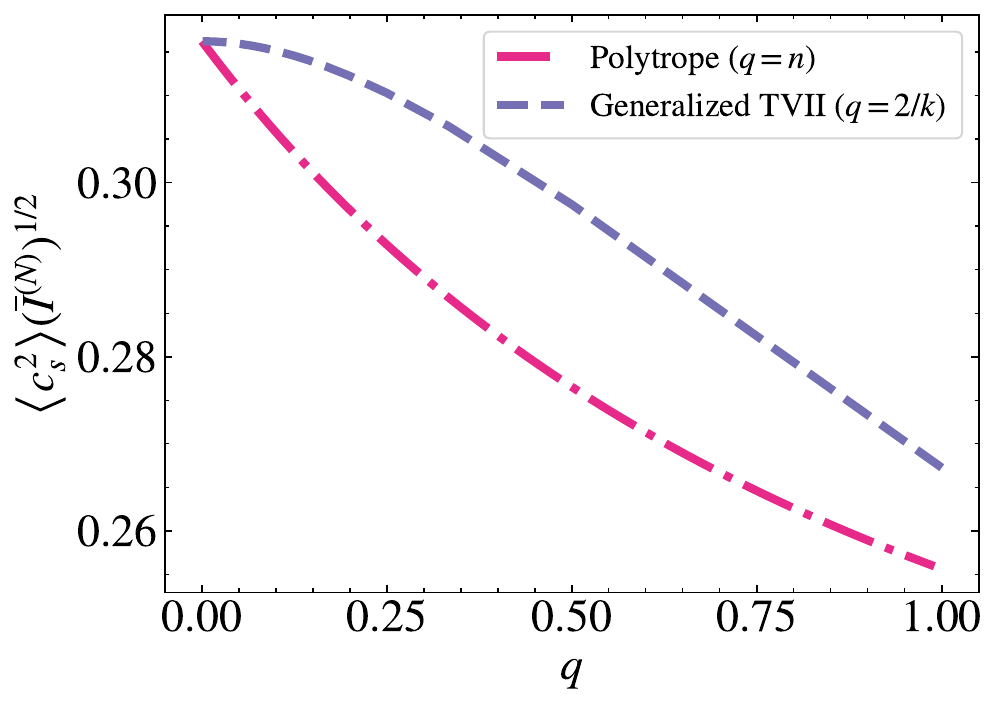}
   \caption{$\avgcs\bar{I}^{1/2}$ as a function of $q=n$ for a polytropic EoS model (dot-dashed pink) and $2/k$ for a generalized Tolman VII model (dashed cyan). 
   Observe that for $q \in [0,1]$, the quantity $\avgcs\bar{I}^{1/2}$ varies little, thus exhibiting an approximate EoS universality. 
   }
   \label{fig:newtrelations_I}
\end{figure}

By substituting the density profile, $\rho(r)$, presented in Eq. \eqref{eq:genTVIIN}, into the integral shown in Eq.~\eqref{eq:Newt_I}, we find that
\begin{equation}\label{eq:INTVIIa}
  \bar{I}^{(N,T)} = \frac{8\pi}{3}\!\! \int_0^R \!\!\! r^4 \rho_c \left[1 - \left(\frac{r}{R}\right)^k \right]  dr = \frac{8\pi }{3}  R^5 \rho_c  \left(\frac{1}{5} - \frac{1}{k+5} \right),
\end{equation}
where the superscript $(N,T)$ indicates we are in the Newtonian regime and using the generalized Tolman VII solution. 
By referring to Eq. \eqref{eq:betaCT7N}, we can establish the relation $4 \pi \rho_c R^2 = {3 C (k+3)}/{k}$. Substituting this relation into Eq.~\eqref{eq:INTVIIa} yields
\begin{align}
\label{eq:INTVII}
    \bar{I}^{(N,T)} &=\frac{2}{C^2} \left(1+\frac{3}{k} \right) \left(\frac{1}{5} - \frac{1}{k+5}\right) 
    \nonumber \\
    &=\frac{k^2 (k+3) (k+4)^2}{10 (k+5) \left(k^2+3 k+2\right)^2}\frac{1}{\avgcs\!\!^2},
\end{align}
where we have used the \avgcs -- $C$ relation for the generalized Tolman VII solution in the Newtonian limit. With this expression, we obtain the following limits when $k=2$ and $k\rightarrow \infty$:
\begin{align}
    k = 2\longrightarrow & \quad \bar{I}^{(N,T)} =  \frac{1}{14\avgcs\!\!^2} \longrightarrow \avgcs \approx 0.27 \left( \bar{I}^{(N,T)} \right)^{-\frac{1}{2}},\\
    k \rightarrow \infty \longrightarrow & \quad  \bar{I}^{(N,T)} = \frac{1}{10\avgcs\!\!^2} \longrightarrow \avgcs \approx 0.32 \left( \bar{I}^{(N,T)} \right)^{-\frac{1}{2}}.
\end{align}
As expected, when $k \rightarrow \infty$, which corresponds to the incompressible case, one obtains the same limit as that found using the polytropic EoS when $n=0$. 
Also, similarly to that case, there is no common limit between the incompressible case and the Tolman VII profile. For $q = 2/k \in [0,1]$, the quantity $\avgcs \left(\bar{I}^{(N,T)} \right)^{1/2}$ displays a maximum fractional difference with respect to the EoS median of $\sim 10\%$. This quantity is shown in Fig.~\ref{fig:newtrelations_I} as a function of $q$.

\begin{figure}
\includegraphics[width=0.4\textwidth  ]{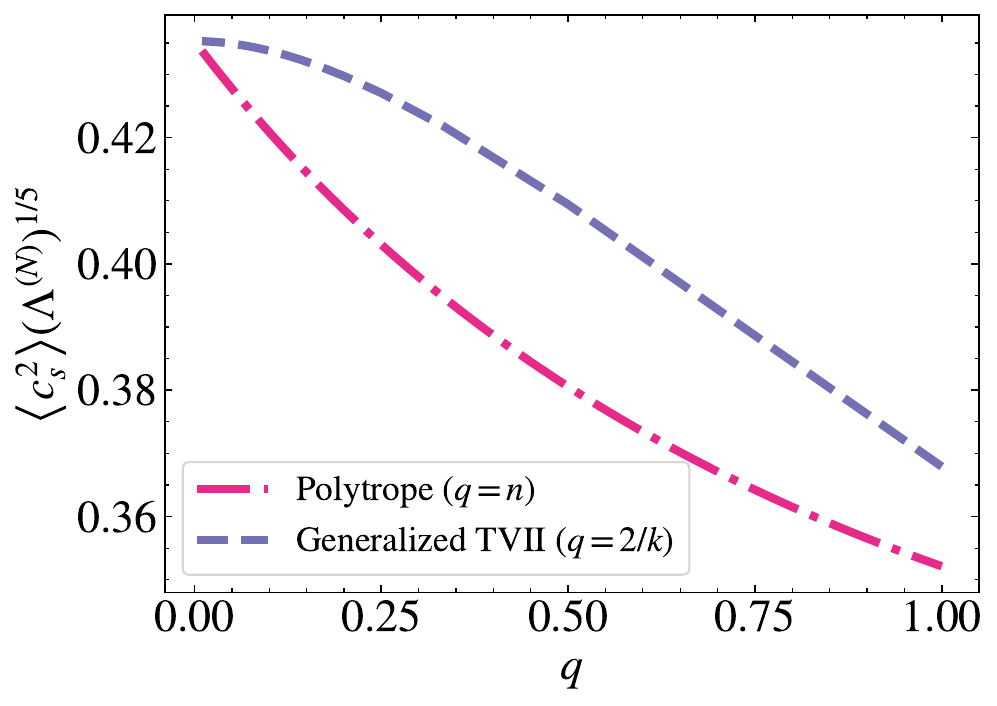}
   \caption{$\avgcs\Lambda^{1/5}$ for two types of solutions to the Newtonian equations of structure, the polytropic EoS (in pink) and the generalized Tolman VII solution (in dashed cyan). We use the polytropic index $n$ for the former, varying it in the interval that is relevant for NSs, and for the latter, we use the quantity $2/k$ that varies between the incompressible case, when $2/k = 0$, and Tolman VII, when $2/k = 1$. }
   \label{fig:newtrelations_lambda}
\end{figure}

For the tidal deformability, we find that
\begin{align}
    k = 2\longrightarrow & \quad \Lambda^{(N,T)} \approx 0.22 \frac{1}{C^5} \longrightarrow \avgcs \approx 0.37 \left( \Lambda^{(N,T)} \right)^{-\frac{1}{5}}  \\
    k \rightarrow \infty \longrightarrow & \quad  \Lambda^{(N,T)} = \frac{1}{2C^5} \longrightarrow \avgcs \approx 0.44 \left( \Lambda^{(N,T)} \right)^{-\frac{1}{5}}.
\end{align}
For $q = 2/k \in [0,1]$, the quantity $\avgcs \left(\Lambda^{(N,T}\right)^{1/5}$ displays a maximum fractional difference with respect to the EoS median of $\sim 10\%$. This quantity is also shown in Fig. \ref{fig:newtrelations_lambda} as a function of $q$.

\subsubsection{Comparing solutions}

The $\avgcs\! - \bar{I}/\Lambda$ relations present a stronger dependence on the EoS than the $\avgcs\! - C$ relation, at least in the Newtonian regime. The variation in the $\avgcs\! - \bar{I}/\Lambda$ relations is depicted in Figs. \ref{fig:newtrelations_I} and \ref{fig:newtrelations_lambda}, respectively, with an EoS-sensitivity of $\sim 14\%$ for a polytropic EoS with $n\in[0,1]$ and $\sim 10\%$ for the generalized Tolman VII solution with $q=2/k \in [0,1]$. To put these numbers in perspective, we can compare them to the EoS-sensitivity of the (Newtonian) $\avgcs\! - C$ relation, $3\%$ for polytropes and $6\%$ for the generalized Tolman VII solution. 

The parallels between the $\avgcs\! - \bar{I}$ and $\avgcs\! - \Lambda$ relations are noteworthy, although unsurprising, as these observables are well known for having a strong universal relation both in GR and in the Newtonian limit \cite{Yagi:2013awa}. Given the known quasi-universal relations $I$-Love-$C$ in GR \cite{Hinderer:2009ca,Postnikov:2010yn}, one might wonder whether the influence of incorporating GR corrections, as observed in the $\avgcs\! - C$ relation, will similarly manifest in the moment of inertia and tidal deformability. We tackle this question next.

\subsection{Post-Minkowskian Expansion}\label{subsec:PMexp}

To assess whether increasing GR contributions will lead to a reduction in the universality for both $\avgcs\! - \bar{I}/\Lambda$ relations, we perform a post-Minkowskian expansion. This enables us to derive GR contributions in an order-by-order manner, facilitating a more precise assessment of the impact of each contribution. In this section, this will be done for the generalized Tolman VII models alone, defined in Eq. \eqref{eq:genTVIIR}. 

The relativistic moment of inertia can be derived under the context of slow rotation, as originally outlined by Hartle \cite{Hartle:1967he}:
\begin{equation} \label{eq:IR}
    I= \frac{8\pi R^5}{3} \int_0^1 dx\, x^4 \frac{(\epsilon + p)}{c^2} e^{(\lambda - \nu)/2} \bar{\omega}.
\end{equation}
Here, $g_{tt} = - e^{\nu}$ and $g_{rr} = e^{\lambda}$ are metric coefficients describing the unperturbed (i.e., non-rotating) spacetime, while $g_{t\phi} = - \Omega (1 - \bar{\omega}) r^2 \sin^2 \theta $ describes the perturbation due to rotation (to first order in the angular velocity $\Omega$). The function $\bar{\omega}(x)$, with $x \equiv r/R$, obeys
\begin{equation}\label{eq:omegabareq}
    \frac{1}{x^4} \frac{d}{dx} \left(\,x^4 j \frac{d\bar{\omega}}{dx} \right) + \frac{4}{x} \frac{dj}{dx} \bar{\omega} = 0,
\end{equation}
with $j \equiv e^{-(\nu+\lambda)/2}$, along with the boundary conditions $d\bar{\omega}/dx|_{x=0} = 0$ and $\lim_{x \to \infty} \bar{\omega}(x) = 1$. 

The function $\lambda(x)$, for the generalized Tolman VII models, can be directly derived from Eq.~(\ref{eq:muTVII}) as $\lambda(x) = - \log (1 - \beta \mu / x)$. For the metric potentials $\nu$ and $\bar{\omega}$, we employ a power series approximation analogous to Eq. \eqref{eq:psexpansionC}:
\begin{equation}
    \nu (x; C) = \sum_{i=0}^N \nu_i (x) C^i, \qquad
    \bar{\omega} (x; C) = \sum_{i=0}^N \bar{\omega}_i (x) C^i.
\end{equation}
We can now solve Eq. \eqref{eq:omegabareq} and the corresponding equation for $\nu$, 
\begin{equation}
\frac{d\nu}{dx} = - \frac{2 \avgcs}{\bar{\epsilon} + \avgcs \bar{p}} \frac{d\bar{p}}{dx}\,,    
\end{equation}
analytically, order by order in $C$, which allows the relativistic moment of inertia \eqref{eq:IR} to be computed in successive post-Minkowskian approximations. In the Newtonian limit (i.e.~when $N=0$), we recover Eq. \eqref{eq:INTVII} for the dimensionless quantity $\bar{I}$. 

\begin{center}
\begin{figure}
\includegraphics[width=0.5\textwidth ]{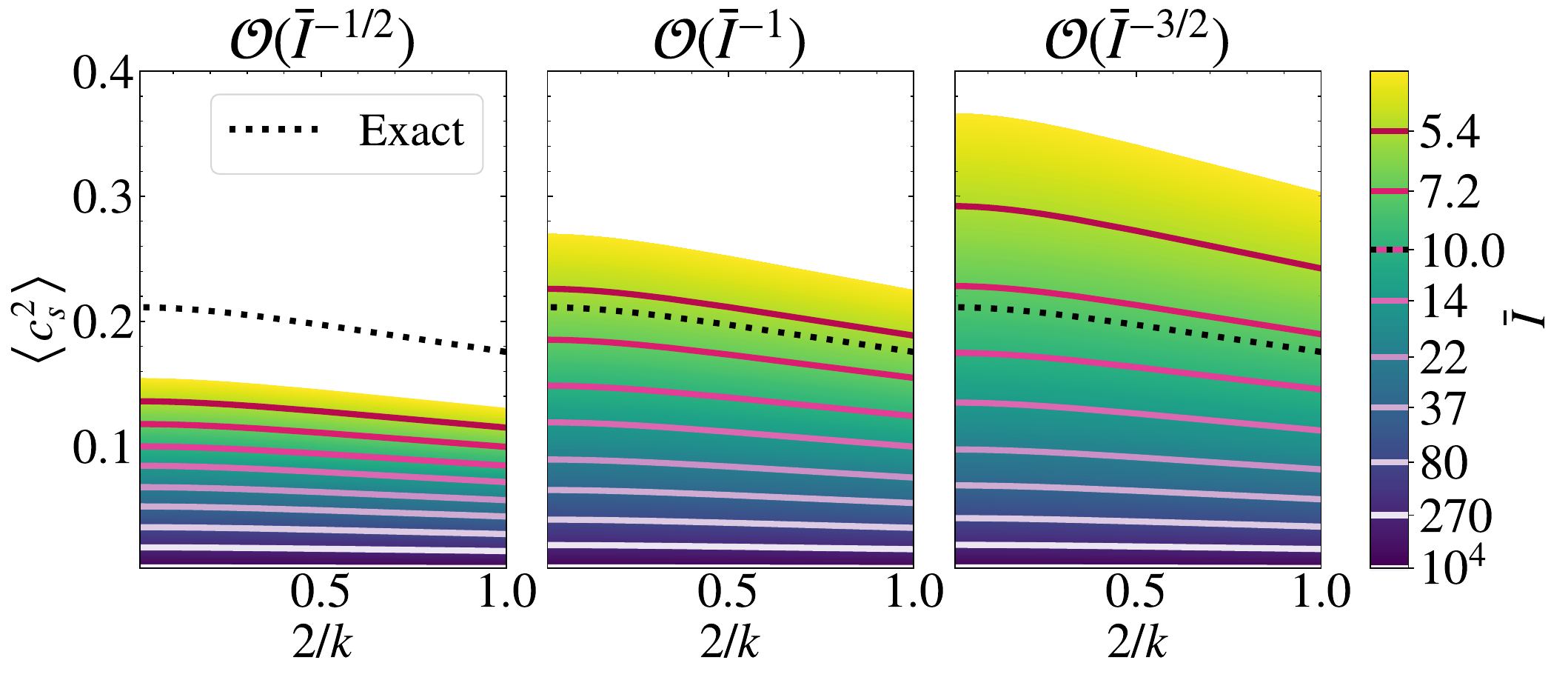}
   \caption{The variation of $\avgcs\! - \bar{I}$ in relation to the post-Minkowskian expansion for the generalized Tolman VII solution, considering values of $k \in [2,\infty)$ along the bottom axis and representing different moment of inertia levels with distinct colors. The expansion analysis starts from the first-order term and progressively incorporates higher-order terms, up to the third order, proceeding from left to right. The black dotted line indicates the numerical solution in full GR for a NS with $\bar{I} = 10$ for varying values of $k$. The minimum value for the moment of inertia is determined by the point where $\avgcs$ becomes greater than 1 for the GR solution.}
   \label{fig:PMGTVIIIcs}
\end{figure}
\end{center}

The relation $\Bar{I} - \avgcs$ can be found using Eq. \eqref{eq:PMGTVIICcs}. As our interest lies in the inverse relation, \avgcs as a function of $\Bar{I}$, we perform a series expansion of the moment of inertia corresponding to an expansion of \avgcs about 0. Since in the Newtonian limit we have that $\avgcs \propto \bar{I}^{-1/2}$, we expand around $\bar{I}$ approaching infinity, or, equivalently, $\bar{I}^{-1/2} \rightarrow 0$. The resulting series converges for $\bar{I}^{-1/2} \lesssim 0.49$, or $\bar{I} \gtrsim 4.2$ (with the radius of convergence being roughly independent of $k$). The effects of this expansion in powers of $\bar{I}^{-1/2}$ on the \avgcs -- $\bar{I}$ relation is depicted in Fig. \ref{fig:PMGTVIIIcs}, where we show values of $\Bar{I}$ for which $\avgcs<1$ in GR. Additionally, we show the full GR curve for a NS with $\bar{I}=10$, which agrees with results displayed in \cite{Silva:2020acr}. Observe that, as we consider higher-order post-Minkowskian orders, $\avgcs$ approaches the full GR value. We expect that the correct solution is obtained when $N\rightarrow \infty$, for values of $\bar{I}^{-1/2}$ within the radius of convergence of the series. 

Importantly, we observe that the increasing post-Minkowskian orders seem to have minimal effects on the EoS-dependence of the \avgcs -- $\bar{I}$ relation.
For instance, for the fiducial dimensionless moment of inertia $\bar{I} = 10$, and considering $q = 2/k \in [0,1]$, the quantity \avgcs displays a maximum fractional difference with respect to the EoS median of $10\%$ to $11\%$ to all orders.
Thus, the degree of universality observed in the Newtonian limit remains consistent as we move to higher post-Minkowskian orders. 
This is to be compared to the effect of post-Minkowskian corrections to the \avgcs -- $C$ relation, which tends to deteriorate the EoS quasi-universality, as discussed earlier. 

Now let us consider the tidal deformability. This quantity is obtained from the generalization of Eq.~(\ref{eq:HeqN}) into
\begin{align}
    &\frac{d^2 h_2}{dr^2} + \frac{dh_2}{dr} \left\{ \frac{2}{r} + e^\lambda \left[ \frac{2Gm}{r^2c^2}- \frac{4\pi Gr}{c^4}(\epsilon - p)\right]\right\} \nonumber \\
    &+h_2 \left[ -\frac{6 e^\lambda}{r^2} + \frac{4\pi G e^\lambda}{c^4} \left(5\epsilon + 9p + \frac{\epsilon + p}{dp/d\epsilon}\right) - \left( \frac{d\nu}{dr}\right)^2 \right]=0, 
\end{align}
where $h_2(r) = - e^{-\nu} \delta g_{tt} $  represents the perturbation to the $(t,t)$ component of the metric, in terms of which all other metric potentials can be obtained.  
The relativistic $l=2$ Love number can be computed from Eq.~(50) of Ref.~\cite{Damour:2009vw}, recalling that $y \equiv R h_2'(R)/h_2(R)$. Furthermore, we again define the dimensionless tidal deformability through Eq.~(\ref{eq:dimensionless}).

Despite $h_2(r)$ having an analytic solution in terms of hypergeometric functions for the generalized Tolman VII profiles in the Newtonian limit, exact analytic solutions could not be found to higher post-Minkowskian orders. Still, to access the impact of post-Minkowskian corrections, we introduce the power series approximation
\begin{equation}\label{eq:Hexpansion}
    h_2(x;C) = \sum_{i = 0}^N h_{2,i}(x) C^i
\end{equation}
where $x = r/R$ and $h_{2,i}(x)$ is calculated numerically.

\begin{center}
\begin{figure}
\includegraphics[width=0.5\textwidth ]{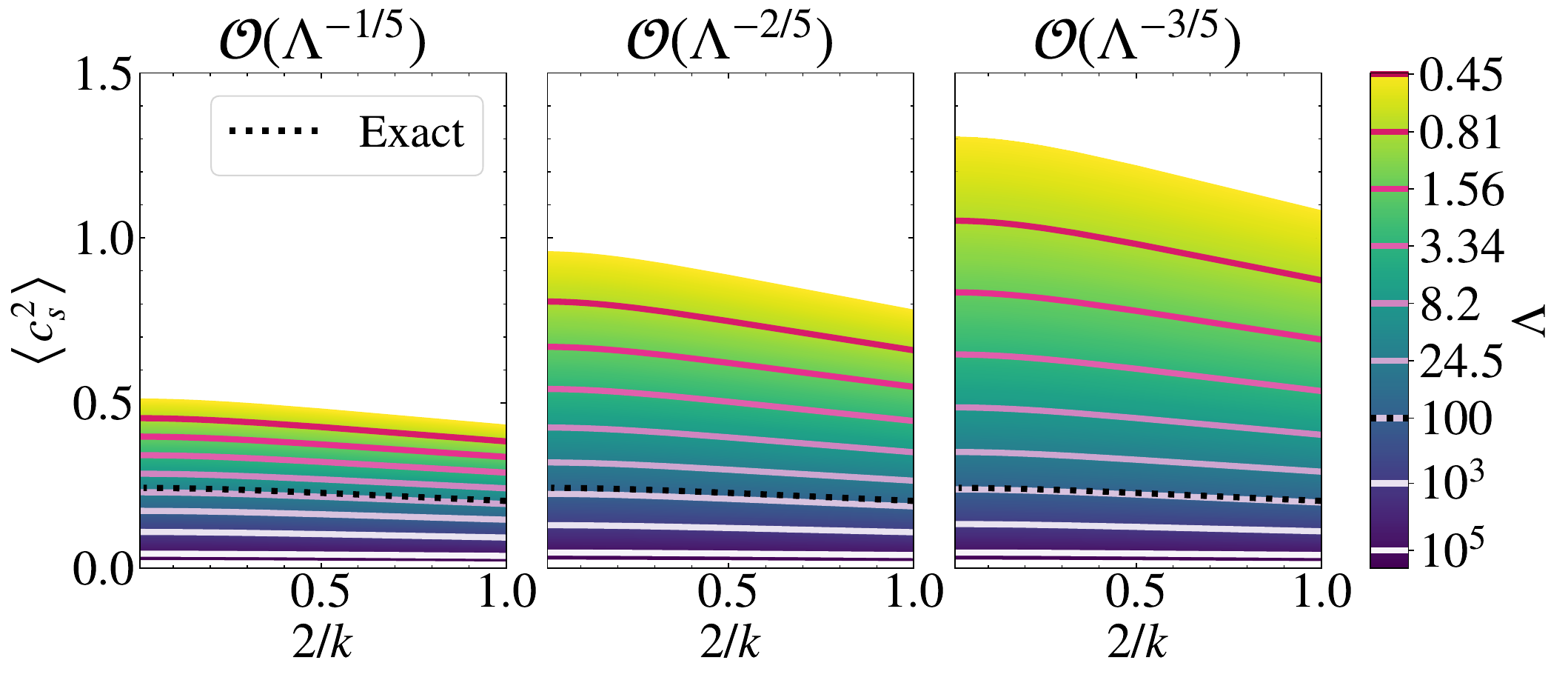}
   \caption{The variation of $\avgcs \! - \Lambda$ in relation to the post-Minkowskian expansion for the generalized Tolman VII solution, considering values of $k \in [2,\infty)$ along the bottom axis and representing different tidal deformability levels with distinct colors. The expansion analysis starts from the first-order term and progressively incorporates higher-order terms, up to the third order, proceeding from left to right. The black dotted lines indicate the numerical solution in full GR for a NS with $\Lambda = 100$ for varying values of $k$.}
   \label{fig:PMGTVIIlambdacs}
\end{figure}
\end{center}

The relation between the dimensionless tidal deformability and \avgcs can be found by rewriting $C$ as a function of \avgcs using Eq. \eqref{eq:PMGTVIICcs}. As we are interested in \avgcs as a function of $\Lambda$, we perform a series expansion of the tidal deformability in powers of \avgcs, as done previously for the moment of inertia. Since in the Newtonian limit we have that $\avgcs \propto \Lambda^{-1/5}$, we expand around $\Lambda$ approaching infinity, or, equivalently, $\Lambda^{-1/5} \rightarrow 0$.  This expansion, up to the third order, is displayed in Fig.~\ref{fig:PMGTVIIlambdacs}, where we show values of $\Lambda$ for which $\avgcs\! <1$ in GR. Additionally, we show the numerical solution for a NS with $\Lambda = 100$, which agrees with measurements for a typical NS \cite{LIGOScientific:2018cki}. Observe again that, as we include higher post-Minkowskian orders, $\avgcs$ approaches the full GR result. We expect that the correct solution is obtained when $N \rightarrow \infty$, for values of $\bar{\Lambda}^{-1/5}$ within the radius of convergence of the series solution.

Importantly, we observe that, for the values for which the series is convergent, the increasing post-Minkowskian orders seem again to have minimal effects on the EoS-dependence of the $\avgcs \! - \Lambda$ relation. For the fiducial tidal deformability $\Lambda = 100$, and considering $q = 2/k \in [0,1]$, the quantity \avgcs displays a maximum fractional difference with respect to the EoS median of $10\%$ to $11\%$ to all orders. 
Thus, the degree of universality observed in the Newtonian limit remains consistent as we move to higher post-Minkowskian orders.

\section{Conclusions} \label{sec:conclusions}

We have here explored the underlying physics responsible for the approximately universal relations between the ratio of the central pressure to the central energy density, $p_c/\epsilon_c$, and dimensionless astrophysical observables, namely, the compactness $C$, the tidal deformability $\Lambda$ and the moment of inertia $\bar{I}$ of NSs~\cite{Saes:2021fzr}. 
The quantity $p_c/\epsilon_c$ was reinterpreted as the squared speed of sound averaged over the range of energy densities present inside the star, $\avgcs\!$. 
The averaged speed of sound squared has a natural interpretation as a measure of the mean stiffness of the EoS up to the central energy density of a given NS.

We first studied the $\avgcs\! - C$ relation using analytic solutions of the TOV equation for an incompressible fluid, a Tolman VII profile, and the Buchdahl fluid. We found that, albeit quite different from each other, these EoSs led to $\avgcs\! - C$ relations with the same Newtonian limit, $\avgcs = C/2$. This prompted further study of the Newtonian limit and its relativistic corrections, which we performed for 2 one-parameter families of EoSs that interpolate between the (Newtonian limit of the) models considered previously: polytropic EoSs and a generalized Tolman VII profile. For the range of parameters most relevant for the description of NSs, the $\avgcs\! - C$ relation was shown to be remarkably flat (with respect to EoS variation) in the Newtonian limit. This insensitivity to the EoS then deteriorated as relativistic corrections were included. For instance, for polytropic EoSs with $n\in[0,1]$, the maximal fractional difference of the $\avgcs\! - C$ relation with respect to the EoS median increased from $3\%$ at $\mathcal{O}(C)$ (Newtonian limit), to $9\%$ at $\mathcal{O}(C^3)$ and 17\% in full GR.

When contrasting realistic EoSs to the one-parameter families considered previously, we found that the latter provides reasonable approximants to the former, as long as the compactness of the NS is sufficiently large ($C\gtrsim 0.1$). Thus, the EoS insensitivity of the $\avgcs\! - C$ relation, for realistic EoSs, can be linked to EoSs being relatively stiff throughout a large portion of the NS for sufficiently large values of the stellar compactness.

In contrast to the case of the $\avgcs\! - C$ relations, we found that the approximate universality of the $\avgcs\! - \bar{I}/\Lambda$ relations is not enhanced in the Newtonian limit. Indeed, for the generalized Tolman VII profile with $k \in [2,\infty)$, we showed that the EoS dependence remains at the $10-11\%$ level to all orders in compactness. This is also the case when using realistic EoSs, as long as the NS is sufficiently compact.

Our work suggests a few avenues for future work. One possibility would be to study potential approximately universal relations between \avgcs and the quadrupole and higher-order multipole moments of rotating NSs. This could be done by considering the slow-rotation expansion~\cite{Hartle:1967he}, and constructing equilibrium sequences of stars with fixed rotational periods. Such a study would also benefit from the post-Minkowskian analysis we laid out in this work. 

\acknowledgments
N. Y. acknowledges support from NSF Award PHY-220765. R. M. acknowledges support from the National Council for Scientific and Technological Development (CNPq), and by the Carlos Chagas Filho Research Support Foundation (FAPERJ). We thank Jaki Noronha-Hostler for helpful discussions about this paper.

\appendix

\section{Equations of structure and analytical solutions}
\label{ap:Equationsofstructure}

\subsection{General Relativity}
\label{subap:GeneralRelativity}
 The relativistic equation of hydrostatic equilibrium is
\begin{align} 
\label{eq:TOVeq}
\frac{dp}{dr} = - \frac{G}{c^2}(\epsilon+p)\frac{m+4\pi r^3 p/c^2}{r\left(r-\frac{2Gm}{rc^2}\right)}\,,
\end{align}
where $m$ is the enclosed mass, $p$ is the internal pressure, $\epsilon$ is the energy density, and $r$ is the radial coordinate. The enclosed mass satisfies the equation:
\begin{align}
\label{eq:massDE}
\frac{dm}{dr}= 4\pi r^2 \frac{\epsilon}{c^2}\,.
\end{align}
These two equations form a closed system once an EoS is prescribed and their solution requires boundary conditions. As usual, we impose regularity at the center of the star, enforcing $m(r=0)=0$, while $p(r=0) = p_c$, where $p_c$ is the central pressure. With these conditions, the radius of the star $R$ is defined as the radial coordinate at which the pressure vanishes, $p(r=R)=0$, while the stellar mass is the enclosed mass evaluated at this radial coordinate $M = m(r=R)$.

\subsection{Newtonian Limit}
\label{subap:Newtonian}

Using that in the Newtonian limit $\epsilon = \rho c^2$, where $\rho$ is the rest mass density, Eqs.~\eqref{eq:TOVeq} and~\eqref{eq:massDE} reduce to 
\begin{align}
    \frac{dp}{dr} = - \rho \frac{m}{r^2},\qquad
    \frac{dm}{dr} = \rho \frac{Gm}{r^2}.
    \end{align}
Just as in the relativistic case, these equations must be closed by choosing an EoS and prescribing boundary conditions. 

\subsection{Analytical solutions to the TOV equations}
\label{subap:AnalyticalSolutions}

In this section we derive expressions \eqref{eq:meancs2incompressible}, \eqref{eq:meancs2tolman}, and \eqref{eq:meancs2buchdal} for the $\avgcs$--$C$ relation for an incompressible fluid, a Tolman VII density profile and the Buchdahl EoS, respectively.

\subsubsection{Incompressible fluid}

A constant-density profile
\begin{align}
\epsilon = \epsilon_c = \text{constant}
\end{align}
results in the following analytic solutions to the TOV \eqref{eq:TOVeq} and the enclosed mass \eqref{eq:massDE} equations
\begin{align}
p(r) & = \frac{\epsilon_c \left( R\sqrt{1-2 C}-\sqrt{R^2-2 C r^2} \right)}{\sqrt{R^2-2 C r^2}-3 R \sqrt{1-2 C} }, \quad m(r)  =  \frac{4}{3} \pi  r^3 \frac{\epsilon_c}{c^2},
\end{align}
where the central energy density $\epsilon_c$ can be expressed in terms of the stellar mass and radius by
\begin{align}
    \epsilon_c = \frac{3 M c^2}{4 \pi  R^3}.
\end{align}
Evaluating $p(r=0)$, we obtain 
\begin{align}
   p_c =\epsilon_c \frac{1 -\sqrt{1-2 C}  }{3 \sqrt{1-2 C}-1},
\end{align}
which leads to the simple analytic relation
\begin{align}
\avgcs  = \frac{p_c}{\epsilon_c}  & =\frac{1 - \sqrt{1 - 2 C}}{3\sqrt{1-2C} - 1}. 
\end{align}

\subsubsection{Tolman VII fluid}

The Tolman VII solution~\cite{Tolman:1939jz} is characterized by the density profile
\begin{align}
      \epsilon(r)  & = \epsilon_c \left[1 - \left(\frac{r}{R}\right)^2\right].
\end{align}
To simplify the resulting analytic expressions presented in \cite{Jiang:2020uvb}, we introduce the dimensionless variable $x = r/R$. The mass aspect function and pressure can be expressed as follows:
\begin{align}
\label{eq:m-Tol}
m(x)  &=\frac{  4 \pi \epsilon_c}{R^3}\left(\frac{x^3}{3} - \frac{x^5}{5}\right),\\
p (x)&=\frac{c^4}{4 \pi  G R^2}\left[ \sqrt{3 C e^{- \lambda}} \tan\phi- \frac{C}{2} (5 - 3 x^2)\right],
\end{align}
where $\epsilon_c$ can be expressed in terms of $R$ and the stellar mass $M$, or, alternatively, the compactness $C$, as
\begin{align}
\epsilon_c = \frac{15 M c^2}{8 \pi  R^3} = \frac{15 c^4 C}{8 \pi G R^2}.
\end{align} 
The functions $\lambda(x)$ and $\phi(x)$ are given by
\begin{eqnarray}
e^{-\lambda(x)} &=& 1- C x^2 \left(5 - 3 x^2\right),\\
\phi (x) &=&k- \frac{1}{2} \log\left(x^2 - \frac{5}{6} +\sqrt{\frac{e^{-\lambda}}{3 C}}\right),
\end{eqnarray}
where
\begin{align}
\label{eq:constTolmanGR}
    k =\arctan \sqrt{\frac{C}{3 (1-2 C)}} + \frac{1}{2}\log\left(\frac{1}{6} + \sqrt{\frac{1 - 2C}{3 C}}\right).
\end{align} 
At the center of the star ($x=0$), we have that $\lambda = 0$ and
\begin{align}
\label{eq:pctolmangr}
p_c= \frac{\epsilon_c}{15}  \left(\frac{2 \sqrt{3} \tan \phi_c}{\sqrt{C}}-5\right),
\end{align}
where 
\begin{equation}
\label{eq:phi_c}
    \phi_c=\phi(0)=k-\frac{1}{2} \log \left(1/\sqrt{3C}-5/6\right).    
\end{equation}

All of the above allows us to easily calculate the averaged speed of sound squared analytically, namely
\begin{align}
\avgcs= \frac{1}{15}  \left(\frac{2 \sqrt{3} \tan \phi_c}{\sqrt{C}}-5\right).
\end{align}

\subsubsection{Buchdahl fluid}

The Buchdahl EoS \cite{1967ApJ...147..310B} is given by
\begin{align}
\epsilon(p) = 12(p_* p)^{1/2} - 5p,
\end{align}
where $p_*$ is a constant. With this EoS, one can find an analytic solution to the TOV equation, which, as presented in~\cite{Lattimer:2000nx}, is
\begin{align}
\label{eq:solbuchdal}
p(r) = \frac{c^4}{8\pi G}A^2u^2(1-2 C)(1-C+u)^{-2},
\end{align}
where 
\begin{align}
r &= r^\prime(1-C+u)(1-2C)^{-1},\\
u &= C(Ar^\prime)^{-1}\sin Ar^\prime,\\
A^2 &= 288\pi p_*Gc^{-4}(1-2C)^{-1}.
\label{eq:buch2}
\end{align}
At the center of the star, $r=r^\prime = 0$ and $u=C$; thus, the central pressure can be expressed as
\begin{align}
\label{eq:pcbuchdal}
p_c = 36 p_* C^2,
\end{align}
and, from Eq. \eqref{eq:buchdaldef}, the central energy density is
\begin{align}
\label{eq:ecbuchdal}
\epsilon_c =  72 p_* C \left(1-\frac{5C}{2}\right).
\end{align}

With all of the above, we can now compute the averaged speed of sound square as
\begin{align}
    \avgcs = \frac{C}{2-5 C}.
\end{align}

\section{Post-Minkowskian expansion of $C - \avgcs$ for polytropic EoSs}
\label{ap:PMexpansions}
Differently from the generalized Tolman VII density profile, for which the post-Minkowskian expansion of the $\avgcs\! - C$ relation can be found analytically, in the case of a generic polytropic EoS, the $\avgcs\! - C$ relation must be obtained numerically. In this Appendix, we sketch the numerical procedure and discuss certain subtleties that render the straightforward computation of high post-Minkowskian coefficients of that expansion numerically challenging.

The compactness $C$ is related to \avgcs by
\begin{equation}\label{eq:Csigma}
    C = (n+1) \sigma \frac{\mu_1^{(P)}(\sigma)}{\xi_1(\sigma)},
\end{equation}
where $\sigma = (1/\avgcs - n)^{-1}$ [cf. Eq.~(\ref{eq:cs2-sigma})], which can be expanded to any desired order. Here, $\mu_1^{(P)} (\sigma) = \mu^{(P)} (\xi_1(\sigma))$, and the dimensionless stellar radius, $\xi = \xi_1$, is determined by
\begin{equation}
    \bar{p} (\xi_1(\sigma); \sigma) = 0.
\end{equation}
Inserting the Taylor expansion of Eq.~\eqref{eq:xi1_expand} for $\xi_1(\sigma)$ in the above equation, we can solve the latter order by order in a post-Minkowskian expansion. For example, at Newtonian order, one obtains 
\begin{equation} \label{eq:xi10}
    p_{(0)}(\xi_{1,0}) = 0,    
\end{equation}
which implicitly defines $\xi_{1,0}$. The coefficient $\xi_{1,i}$ with $i\geq 1$ then depends on the respective pressure coefficient $p_{(i)}$, as well as on derivatives of the lower-order pressure coefficients ($p_{(j)}$, $j<i$), evaluated at $\xi_{1,0}$. Explicitly, for $i = 1, 2$, one has
\begin{align}
    \xi_{1,1}& = -\frac{p_{(1)}(\xi_{1,0})}{p'_{(0)}(\xi_{1,0})}, \label{eq:xi11} \\
    \xi_{1,2} &= - \frac{p_{(2)}(\xi_{1,0}) + \xi_{1,1} p'_{(1)} (\xi_{1,0}) + \frac{1}{2}\xi_{1,1}^2 p''_{(0)} (\xi_{1,0})}{p'_{(0)}(\xi_{1,0})}.\label{eq:xi12}
\end{align}

The numerical procedure then consists of the following steps:
\begin{itemize}
    \item[(i)] numerically solving the Newtonian equations for $p_{(0)}$ and $\mu_{(0)}^{(P)}$, and extracting the value of $\xi_{1,0}$ such that Eq.~\eqref{eq:xi10} holds, as well as $\mu_{1,0}^{(P)}$; 
    \item [(ii)] numerically solving the equations for $p_{(i)}$, $\mu_{(i)}^{(P)}$, for $i\geq 1$ up to the desired order in the fixed domain $\xi\in [0,\xi_{1,0}]$, and computing the coefficients $\xi_{1,i}$ and $\mu_{1,i}^{(P)}$; 
    \item [(iii)] computing the  $C - \avgcs$ relation from the Taylor-expanded version of Eq.~(\ref{eq:Csigma}).
\end{itemize}
    
Now, it can be shown that, around $\xi = \xi_{1,0}$, the $i$-th pressure coefficient has the leading behavior $p_{(i)} \propto (\xi - \xi_{1,0})^{n-i+1}$. In the range of interest $0\leq n \leq 1$, Eq.~(\ref{eq:xi11}) presents a $0/0$ indeterminacy, and must be computed by taking the limit of that expression as $\xi \to \xi_{1,0}$. This limit is expected to be finite and is easily handled numerically. However, in the same range $0\leq n \leq 1$, Eq.~(\ref{eq:xi12}) displays an apparent divergence in the limit $\xi \to \xi_{1,0}$, since the numerator goes as $(\xi-\xi_{1,0})^{n-1}$ and the denominator goes as $(\xi-\xi_{1,0})^n$. A careful analysis of the differential equation for $p_{(2)}$ shows, however, that the coefficient of the dominant term in the numerator of Eq.~(\ref{eq:xi12}), i.e., $\propto (\xi-\xi_{1,0})^{n-1}$, vanishes exactly, so that the true leading-order behavior of both the numerator and denominator is $(\xi-\xi_{1,0})^{n}$, yielding a finite value for the ratio in the limit $\xi \to \xi_{1,0}$. A similar situation is expected for higher-order coefficients ($\xi_{1,i}$, $i>2$).
However, numerically the cancellation of diverging terms described above is not guaranteed due to the inevitable presence of numerical errors, and a higher-precision calculation may be required.

\bibliography{lib}

\end{document}